\documentclass[a4paper]{report}
\usepackage[utf8]{inputenc}
\usepackage[T1]{fontenc}
\usepackage{RJournal}
\usepackage{amsmath,amssymb,array}
\usepackage{booktabs}

\providecommand{\tightlist}{%
  \setlength{\itemsep}{0pt}\setlength{\parskip}{0pt}}

\usepackage{longtable}

\newlength{\cslhangindent}
\newlength{\csllabelwidth}
\newlength{\cslentryspacingunit} % times entry-spacing
  {}%
  {\par}
\newenvironment{CSLReferences}[2] % #1 hanging-ident, #2 entry spacing
 {% don't indent paragraphs
  \setlength{\parindent}{0pt}
  \ifodd #1
  \let\oldpar\par
  \def\par{\hangindent=\cslhangindent\oldpar}
  \fi
  \setlength{\parskip}{#2\cslentryspacingunit}
 }%
 {}
\usepackage{calc}

\usepackage{subcaption}
\usepackage{tikz}
\usetikzlibrary{positioning}
\usetikzlibrary{shapes.geometric, arrows}

\begin{document}

\sectionhead{Contributed research article}
\volume{15}
\volnumber{3}
\year{2023}
\month{September}
\setcounter{page}{93}

\begin{article}

\title{Statistical Models for Repeated Categorical Ratings: The R Package \pkg{rater}}
\author{by Jeffrey M. Pullin, Lyle C. Gurrin, and Damjan Vukcevic}

\maketitle

\abstract{%
A common problem in many disciplines is the need to assign a set of items into categories or classes with known labels. This is often done by one or more expert raters, or sometimes by an automated process. If these assignments or `ratings' are difficult to make accurately, a common tactic is to repeat them by different raters, or even by the same rater multiple times on different occasions. We present an R package \CRANpkg{rater}, available on CRAN, that implements Bayesian versions of several statistical models for analysis of repeated categorical rating data. Inference is possible for the true underlying (latent) class of each item, as well as the accuracy of each rater. The models are extensions of, and include, the Dawid--Skene model, and we implemented them using the Stan probabilistic programming language. We illustrate the use of \CRANpkg{rater} through a few examples. We also discuss in detail the techniques of marginalisation and conditioning, which are necessary for these models but also apply more generally to other models implemented in Stan.
}

This article was originally published in \emph{The R
Journal} (\url{https://doi.org/10.32614/RJ-2023-064}).  The current
version has largely the same content but with the formatting optimised for PDF.

\hypertarget{sec:intro}{%
\section{Introduction}\label{sec:intro}}

The practice of measuring phenomena by having one or more raters assign to items
one of a set of ratings is common across many fields. For example, in medicine
one or more doctors may classify a diagnostic image as being evidence for one of
several diagnoses, such as types of cancer. This process of rating is often done
by one or more expert raters, but may also be performed by a
large group of non-experts (e.g., in natural language processing
(NLP), where a large number of crowd-sourced raters are used to classify
pieces of text; see Passonneau and Carpenter (2014) and Ipeirotis, Provost, and Wang (2010)) or by a machine
or other automated process (e.g., a laboratory diagnostic test, where `test'
may refer to either the raters and/or the ratings). Other fields where ratings
occur include astronomy, for classifying astronomical images (Smyth et al.\ 1994), and
bioinformatics, for inferring error rates for some bioinformatics algorithms
(Jakobsdottir and Weeks 2007). Indeed, both studies apply the Dawid--Skene model that we
present below and implement in our software package.

The assignment of items to categories, which are variously referred to as gradings,
annotations or labelled data, we will call \emph{ratings}. Our hope is that
these are an accurate reflection of the true characteristics of the items being
rated. However, this is not guaranteed; all raters and rating systems are fallible.
We would expect some disagreement between the raters, especially from non-experts,
and even expert raters may disagree when there are some items that are more difficult
to rate than others. A typical way of dealing with this problem is to obtain multiple
ratings for each item from different raters, some of whom might rate the item
more than once on separate occasions. By using an aggregate index (e.g., by averaging
the ratings) one can hope to reduce bias and variance when estimating both
population frequencies and individual item categories.

Despite the fact that ratings data of this type are common, there are few software
packages that can be used to fit useful statistical models to them.
To address this, we introduce \CRANpkg{rater}, a software package for R (R Core Team 2021)
designed specifically for such data and available on CRAN. Our package provides the
ability to fit the Dawid--Skene model (Dawid and Skene 1979) and several of its extensions.
The goal is to estimate accurately the underlying true class of each item (to the
extent that this is meaningful and well-defined) as well as to quantify and
characterise the accuracy of each rater. Models of these data should account
for the possibility that accuracy differs between the raters and that the errors
they make might be class-specific. The package accepts data in multiple formats
and provides functions for extracting and visualizing parameter estimates.

While our package implements optimisation-based inference, the default mode of
inference is Bayesian, using Markov Chain Monte Carlo (MCMC). We chose a Bayesian
approach for several reasons. The first is because the standard classical approach,
using an EM algorithm, often gives parameter estimates on the boundary of the
parameter space (i.e., probabilities of exactly 0 or 1), which would typically
be implausible. Examples of this can be seen in the original paper by Dawid and Skene (1979),
and also in our own results below that compare MCMC and optimisation-based
inference (e.g., compare \autoref{fig:plot-theta} and
\autoref{fig:plot-theta-fit2}). A second reason is to take advantage of the
convenient and powerful Stan software (Stan Development Team 2021), which made implementing our
package considerably easier. Stan is a probabilistic programming language that
implements MCMC and optimisation algorithms for fitting Bayesian models. The
primary inference algorithm used by Stan is the No U-Turn Sampler (NUTS) variant
of the Hamiltonian Monte Carlo (HMC) sampler, which uses gradient information
to guide exploration of the posterior distribution. A third reason is that it allows
users to easily incorporate any prior information when it is available. In this
context, prior information might be information about the quality of the raters
or specific ratings mistakes that are more common.

We illustrate our methods and software using two examples where the raters are
healthcare practitioners: one with anaesthetists rating the health of patients,
and another with dentists assessing the health of teeth. In both examples, we
are interested in obtaining more accurate health assessments by combining all
of the experts' ratings rather than using the ratings of a single expert,
as well as estimating the accuracy of each expert and characterising the types
of errors they make.

Our paper is organised as follows. \protect\hyperlink{sec:data-formats}{Section 2} describes the
various data formats that can be used with the function implemented in
the package. \protect\hyperlink{sec:models}{Section 4} introduces the Dawid--Skene model and
several of its extensions. \protect\hyperlink{sec:implementation}{Section 5} briefly describes
the implementation and goes into more detail about some of the more
interesting aspects, followed by an exposition of the technique called
Rao--Blackwellization in \protect\hyperlink{sec:rao-blackwellization}{Section 6}, which is
connected with many aspects of the implementation. \protect\hyperlink{sec:usage}{Section 7}
introduces the user interface of \CRANpkg{rater} along with some
examples. \protect\hyperlink{sec:summary}{Section 8} concludes with a discussion of the potential
uses of the model.

\hypertarget{sec:data-formats}{%
\section{Data formats}\label{sec:data-formats}}

There are several different formats in which categorical rating data can
be expressed. We make a distinction between \emph{balanced} and \emph{unbalanced}
data. Balanced data have each item rated at least once by each rater.
Unbalanced data contain information on some items that were not rated by
every rater; stated alternatively, there is at least one rater who does
not rate every item.

\autoref{tab:data-formats-balanced} and \autoref{tab:data-formats-unbalanced}
illustrate three different formats for categorical rating data. Format~(a)
has one row per item-rater rating episode. It requires variables
(data fields) to identify the item, the rater and the rating. This
presentation of data is often referred to as the `long' format. It is
possible for a rater to rate a given item multiple times; each such
rating would have its own row in this format. Format~(b) has one row per
item, and one column per rater, with each cell being a single rating.
Format~(c) aggregates rows (items) that have identical rating patterns
and records a frequency tally. Formats~(b) and~(c) are examples of
`wide' formats. They only make sense for data where each rater rates
each item at most once. Note that the long format will never require
any structural missing values for any of the variables, whereas the wide formats
require the use of a missing value indicator unless they represent data
generated by a balanced design, see \autoref{tab:data-formats-unbalanced}.
One benefit of the grouped format is that it allows a more
computationally efficient implementation of the likelihood function. See
\protect\hyperlink{sec:implementation}{Section 5} for details about how this is implemented in
\CRANpkg{rater}.

\begin{table}
    \caption{Formats for balanced rating data.}
    \centering
    \begin{subtable}[t]{.3\linewidth}
        \centering
        \caption{Long}
        \begin{tabular}{ccc}
            \toprule
            {Item} & {Rater} & {Rating}  \\
            \midrule
            1 & 1 & 3 \\
            1 & 2 & 4 \\
            2 & 1 & 2 \\
            2 & 2 & 2 \\
            3 & 1 & 2 \\
            3 & 2 & 2 \\
        \bottomrule
        \end{tabular}
    \end{subtable}%
    \begin{subtable}[t]{.33\linewidth}
        \centering
        \caption{Wide}
        \begin{tabular}{ccc}
            \toprule
            {Item} & {Rater 1} & {Rater 2} \\
            \midrule
             1 & 3 & 4 \\
             2 & 2 & 2 \\
             3 & 2 & 2 \\
            \bottomrule
        \end{tabular}
    \end{subtable}%
    \begin{subtable}[t]{.33\linewidth}
        \centering
        \caption{Wide (grouped)}
        \begin{tabular}{ccc}
            \toprule
            {Rater 1} & {Rater 2} & {Tally} \\
            \midrule
             3 & 4 & 1 \\
             2 & 2 & 2 \\
            \bottomrule
        \end{tabular}
    \end{subtable}%
    \label{tab:data-formats-balanced}
\end{table}

\begin{table}
    \caption{Formats for unbalanced rating data.
             Missing values are indicated by 'NA'.}
    \centering
    \begin{subtable}[t]{.3\linewidth}
        \centering
        \caption{Long}
        \begin{tabular}{ccc}
            \toprule
            {Item} & {Rater} & {Rating}  \\
            \midrule
            1 & 1 & 3 \\
            1 & 2 & 4 \\
            2 & 1 & 2 \\
            2 & 2 & 2 \\
            3 & 1 & 2 \\
            3 & 2 & 2 \\
            4 & 1 & 3 \\
            5 & 1 & 3 \\
            6 & 2 & 4 \\
        \bottomrule
        \end{tabular}
    \end{subtable}%
    \begin{subtable}[t]{.33\linewidth}
        \centering
        \caption{Wide}
        \begin{tabular}{ccc}
            \toprule
            {Item} & {Rater 1} & {Rater 2} \\
            \midrule
             1 & 3 & 4 \\
             2 & 2 & 2 \\
             3 & 2 & 2 \\
             4 & 3 & NA \\
             5 & 3 & NA \\
             6 & NA & 4 \\
            \bottomrule
        \end{tabular}
    \end{subtable}%
    \begin{subtable}[t]{.33\linewidth}
        \centering
        \caption{Wide (grouped)}
        \begin{tabular}{ccc}
            \toprule
            {Rater 1} & {Rater 2} & {Tally} \\
            \midrule
             3 & 4 & 1 \\
             2 & 2 & 2 \\
             3 & NA & 2 \\
             NA & 4 & 1 \\
            \bottomrule
        \end{tabular}
    \end{subtable}%
    \label{tab:data-formats-unbalanced}
\end{table}

\hypertarget{sec:existing-approaches}{%
\section{Existing approaches}\label{sec:existing-approaches}}

A typical approach for modelling categorical ratings is to posit an unobserved underlying
true class for each item, i.e., we represent the true class as
a latent variable. One of the first such models was developed by
Dawid and Skene (1979). This model is still studied and used to this day, and has
served as a foundation for several extensions (e.g., Paun et al.\ 2018). Our
R package, \CRANpkg{rater}, is the first one specifically designed to provide
inference for the Dawid--Skene model and its variants.

The Python library \pkg{pyanno} (Berkes et al.\ 2011) also implements a Bayesian
version of the Dawid--Skene model in addition to the models described by
Rzhetsky (2009). However, unlike \CRANpkg{rater}, \pkg{pyanno} only
supports parameter estimation via optimisation rather than MCMC; i.e., it will
only compute posterior modes rather than provide samples from the full posterior
distribution. In addition, \pkg{pyanno} does not support variants of the
Dawid--Skene model nor the support for grouped data implemented in
\CRANpkg{rater}.

More broadly, many different so-called `latent class' models have been
developed and implemented in software, for a wide diversity of applications
(e.g., Goodman 1974). A key aspect of the categorical ratings context
is that we assume our categories are known ahead of time, i.e., the possible
values for the latent classes are fixed. The Dawid--Skene model has this
constraint built-in, whereas other latent class models are better tailored to
other scenarios. Nevertheless, many of these models are closely related and
are implemented as R packages.

Of most interest for categorical ratings analysis are the packages
\CRANpkg{poLCA} (Linzer and Lewis 2011), \CRANpkg{BayesLCA} (White and Murphy 2014) and
\CRANpkg{randomLCA} (Beath 2017), all of which are capable of fitting
limited versions of the Dawid--Skene model. We explore the relationship
between different latent class models in more detail in \protect\hyperlink{sec:relationships-existing-models}{Section
4.7}. Briefly: \CRANpkg{randomLCA} can
fit the Dawid--Skene model only when the data uses binary categories,
\CRANpkg{BayesLCA} can only fit the homogeneous Dawid--Skene model, and
\CRANpkg{poLCA} can fit the Dawid--Skene model with an arbitrary number of
categories but only supports wide data (where raters do not make repeated
ratings on items). Neither \CRANpkg{poLCA} nor \CRANpkg{randomLCA} support
fully Bayesian inference with MCMC, and none of the packages support fitting
variants of the Dawid--Skene model (which are available in \CRANpkg{rater}).

\hypertarget{sec:models}{%
\section{Models}\label{sec:models}}

One of the first statistical models proposed for categorical rating data was
that of Dawid and Skene (1979). We describe this model below, extended to include
prior distributions to allow for Bayesian inference. Recently, a number of
direct extensions to the Dawid--Skene model have been proposed. We describe
several of these below, most of which are implemented in \CRANpkg{rater}. For
ease of exposition, unless otherwise stated, all notation in this section will
assume we are working with balanced data and where each item is rated exactly
once by each rater, one exception is that we present the Dawid--Skene model
for arbitrary data at the end of \protect\hyperlink{sec:dawid-skene}{Section 4.1}.

Assume we have \(I\) items (for example, images, people, etc.) and \(J\) raters,
with each item presumed to belong to one of the \(K\) categories: we refer to
this as its `true' category and also as its latent class (since it is unobserved).
Let \(y_{i, j} \in \{1, \dots, K\}\) be the rating for item \(i\) given by
rater \(j\).

\hypertarget{sec:dawid-skene}{%
\subsection{Dawid--Skene model}\label{sec:dawid-skene}}

The model has two sets of parameters:

\begin{itemize}
\item
  \(\pi_k\): the prevalence of category \(k\) in the population from which
  the items are sampled. \(\pi_k \in (0, 1)\) for
  \(k \in \{1, \dots, K\}\), with \(\sum_{k = 1}^K \pi_i = 1\). All classes
  have some non-zero probability of occurring. We collect
  all the \(\pi_k\) parameters together into a vector,
  \(\pi = (\pi_1, \dots, \pi_K)\).
\item
  \(\theta_{j, k, k'}\): the probability that rater \(j\) responds with
  class \(k'\) when rating an item of true class \(k\). Here, \(j \in \{1,  \dots, J\}\) and \(k, k' \in \{1, 2, ..., K\}\). We will refer
  to the \(K  \times K\) matrix \(\theta_{j, k, k'}\) for a given \(j\) as the
  \emph{error matrix} for rater \(j\). We represent the \(k\)th row of this
  matrix as the vector
  \(\theta_{j, k} = \left(\theta_{j, k, 1}, \theta_{j, k, 2},  \dots, \theta_{j, k, K}\right)\): this shows how rater \(j\)
  responds to an item with true class \(k\), by rating it as being in
  class \(1, 2,  \dots, K\), according to the respective probabilities.
\end{itemize}
The model also has the following set of latent variables:
\begin{itemize}
\tightlist
\item
  \(z_i\): the true class of item \(i\). \(z_i \in \{1 ..., K\}\) for \(i \in  \{1, ..., I\}\).
\end{itemize}
Under the Bayesian perspective, the model parameters and latent variables are both
unobserved random variables that define the joint probability
distribution of the model. As such, inference on them is conducted in
the same way. However, we make the distinction here to help clarify
aspects of our implementation later on.

The model is defined by the following distributional assumptions:
\[
\begin{aligned}
z_i &\sim
   \textrm{Categorical}(\pi),
   \quad \forall i \in \{1, \dots, I\}, \\
y_{i, j} \mid z_i &\sim
   \textrm{Categorical}(\theta_{j, z_i}),
   \quad \forall i \in \{1, \dots, I\},
   j \in \{1, \dots, J\}.
\end{aligned}
\]
In words, the rating for item \(i\) given by rater \(j\) will follow the
distribution specified by: the error matrix for rater \(j\), but taking
only the row of that matrix that corresponds to the \textbf{value} of the
latent variable for item \(i\) (the \(z_i\)th row).

We now have a fully specified model, which allows use of
likelihood-based methods. The likelihood function for the observed data
(the \(y_{i,j}\)s) is
\[
  \Pr(y \mid \theta, \pi) =
  \prod_{i = I}^{I} \left(\sum_{k = 1}^{K}
  \left(\pi_{k} \cdot \prod_{j = 1}^{J} \theta_{j, k, y_{i,j}}\right)\right).
  \label{eq:eqnmarginalise}
\]

Note that the unobserved latent variables,
\(z = ({z}_{1}, {z}_{2}, \ldots, {z}_{I})\), do not appear because they
are integrated out (via the sum over the categories \(k\)). Often it is
useful to work with the complete data likelihood where the \(z\) have
specific values, for example if implementing an EM algorithm (such as
described by Dawid and Skene (1979)). The somewhat simpler likelihood function in
that case is
\[
  \Pr(y, z \mid \theta, \pi) =
  \prod_{i = I}^{I} \left(\pi_{z_{i}} \cdot
  \prod_{j = 1}^{J} \theta_{j, z_{i}, y_{i,j}}\right).
\]
In our implementation we use the first version of the likelihood, which
marginalises over \(z\) (see \protect\hyperlink{sec:marginalisation}{Section~5.1}). We do this to avoid
needing to sample from the posterior distribution of \(z\) using Stan,
as the HMC sampling algorithm used by Stan requires gradients with respect to
all parameters and gradients cannot be calculated for discrete
parameters such as \(z\). Alternative Bayesian implementations, perhaps using
Gibbs sampling approaches, could work with the second version of the likelihood
and sample values of \(z\) directly.

To allow for Bayesian inference, we place weakly informative prior
probability distributions on the parameters:
\[
\begin{aligned}
\pi            &\sim \textrm{Dirichlet}(\alpha), \\
\theta_{j, k}  &\sim \textrm{Dirichlet}(\beta_k).
\end{aligned}
\]
The hyper-parameters defining these prior probability distributions are:

\begin{itemize}
\item
  \(\alpha\): a vector of length \(K\) with all elements greater than 0
\item
  \(\beta\): a \(K \times K\) matrix with all elements greater than 0,
  with \(\beta_k\) referring to the \(k\)th row of this matrix.
\end{itemize}

We use the following default values for these hyper-parameters in
\CRANpkg{rater}: \(\alpha_k = 3\) for
\(k \in \{1, \dots, K\}\), and
\[
    \beta_{k, k'} =
    \begin{cases}
           N      p          & \textrm{if } k = k' \\
    \dfrac{N (1 - p)}{K - 1} & \textrm{otherwise}
    \end{cases}
    \quad \forall k, k' \in \{1, \dots, K\}.
\]
Here, \(N\) corresponds to an approximate pseudocount of hypothetical
prior observations and \(p\) an approximate probability of a correct
rating (applied uniformly across all raters and categories). This
affords us the flexibility to centre the prior on a particular assumed
accuracy (via \(p\)) as well as tune how informative the prior will be
(via \(N\)). In \CRANpkg{rater} we set the default values to be \(N = 8\) and \(p = 0.6\),
reflecting a weakly held belief that the raters are have a better than coin-flip
chance of choosing the correct class. This should be suitable for many datasets,
where the number of categories is small, for example, less than ten.
This default would, however, be optimistic if the number of categories is very
large, for example, one hundred. In that case it would make sense for the user
to specify a more realistic prior (see \protect\hyperlink{sec:different-models-priors}{Section 7.5}).
A derivation of the default prior specification
is shown in \protect\hyperlink{sec:hyper-parameters}{Section 4.2}.

We can also write the Dawid--Skene model in notation that does not
assume the data are balanced. Let \(I\), \(J\), \(K\) be as above. Let \(N\) be
the total number of ratings in the dataset (previously we had the
special case \(N = I \cdot J\)). Define the following quantities relating
to the \(n\)th rating: it was performed by rater \(j_n\), who rated item
\(i_n\), and the rating itself is \(y_n\). We can now write the Dawid--Skene
model as:
\[
\begin{aligned}
z_i &\sim
    \textrm{Categorical}(\pi),
    \quad \forall i \in \{1, \dots, I\}, \\
y_n \mid z_{i_n} &\sim
    \textrm{Categorical}
    (\theta_{j_n, z_{i_n}}),
    \quad \forall n \in \{1, \dots, N\}.
\end{aligned}
\]

\hypertarget{sec:hyper-parameters}{%
\subsection{Hyper-parameters for the Dawid--Skene model}\label{sec:hyper-parameters}}

\hypertarget{hyper-parameters-for-the-error-matrices}{%
\subsubsection{Hyper-parameters for the error matrices}\label{hyper-parameters-for-the-error-matrices}}

The values we proposed for the \(\beta\) hyper-parameters are designed to
be flexible and have an approximate intuitive interpretation. The
derivation is as follows.

First, consider the variant of the model where the true latent class
(\(z\)) of each item is known; this model is described under `Case 1.~True
responses are available' by Dawid and Skene (1979). Under this model, we can ignore
\(\pi\) because the distribution of \(y\) depends only on \(\theta\). It is a
categorical distribution with parameters given by specific rows of the
error matrices (as determined by \(z\), which is known) rather than being
a sum over all possible rows (according to the latent class
distribution).

Under this model, the Dirichlet prior is conjugate; we obtain a
Dirichlet posterior for each row of each rater's error matrix. Consider
an arbitrary rater \(j\) and an arbitrary latent class \(k'\). Let
\(c = (c_1, c_2, \dots, c_K)\) be the number of times this rater rates an
item as being in each of the \(K\) categories when it is of true class
\(k'\). Let \(n_{jk'} = \sum_{k = 1}^K c_k\) be the total number of such
ratings. Also, recall that \(\theta_{j, k'}\) refers to the vector of
probabilities from the \(k'\)th row of the error matrix for rater \(j\). We
have that
\[
c \sim \textrm{Multinomial}\left(n_{jk'}, \theta_{j, k'}\right),
\]
giving the posterior
\[
\theta_{j, k'} \mid c
\sim \textrm{Dirichlet}(\beta_{k'} + c).
\]
Under this model, the hyper-parameter vector \(\beta_{k'}\) has the same
influence on the posterior as the vector of counts \(c\). We can therefore
interpret the hyper-parameters as `pseudocounts'. This gives a
convenient way to understand and set the values of the hyper-parameters.
Our choices were as follows:

\begin{itemize}
\item
  We constrain their sum to be a given value
  \(N = \sum_{k = 1}^K \beta_{k',  k}\), which we interpret as a pseudo-sample size (of hypothetical
  ratings from rater \(j\) of items with true class \(k'\)).
\item
  We then set \(\beta_{k', k'}\) to a desired value so that it reflects
  a specific assumed accuracy. In particular, the prior mean of
  \(\theta_{k',  k'}\), the probability of a correct rating, will be
  \(\beta_{k', k'} / N\). Let this be equal to \(p\), an assumed average
  prior accuracy.
\item
  Finally, in the absence of any information about the different
  categories, it is reasonable to treat them as exchangeable and thus
  assume that errors in the ratings are on average uniform across
  categories. This implies that the values of all of the other
  hyper-parameters (\(\beta_{k',  k}\) where \(k \neq k'\)) are equal and need to be set to a value
  to meet the constraint that the sum of the hyper-parameters in the
  vector \(\beta_{k'}\) is \(N\).
\item
  These choices imply the hyper-parameter values described in
  \protect\hyperlink{sec:dawid-skene}{Section 4.1}.
\end{itemize}

In the Dawid--Skene model, the latent classes (\(z\)) are of course not
known. Therefore, we do not have a direct interpretation of the \(\beta\)
hyper-parameters as pseudocounts. However, we can treat them
approximately as such.

We chose \(N = 8\) and \(p = 0.6\) as default values based on obtaining
reasonably good performance in simulations across a diverse range of
scenarios (data not shown).

\hypertarget{relationship-to-previous-work}{%
\subsubsection{Relationship to previous work}\label{relationship-to-previous-work}}

The \emph{Stan User's Guide} (Stan Development Team 2021, sec.\ 7.4) suggests the following choice of
values for \(\beta\):
\[
  \beta_{k', k} =
    \begin{cases}
      2.5 \times K & \textrm{if } k' = k \\
      1 & \textrm{otherwise}
    \end{cases},
  \quad \forall k', k \in \{1, \dots, K\}.
\]
It is interesting to compare this to our suggested values, above. By
equating the two, we can re-write the Stan choice in terms of \(p\) and
\(N\):
\[
\begin{aligned}
  p &= 2.5 \times K / N  \\
  N &= 3.5 \times K - 1.
\end{aligned}
\]
We see that the Stan default is to use an increasingly informative prior
(\(N\) gets larger) as the number of categories (\(K\)) increases. The
assumed average prior accuracy also varies based on \(K\). For example:

\begin{itemize}
\item
  For \(K = 2\) categories, \(p = 0.83\) and \(N = 6\).
\item
  For \(K = 5\) categories, \(p = 0.76\) and \(N = 16.5\).
\item
  As \(K\) increases, \(p\) slowly reduces, with a limiting value of
  \(2.5 / 3.5 = 0.71\).
\end{itemize}

Our default values of \(p = 0.6\) and \(N = 8\) give a prior that is less
informative and less optimistic about the accuracy of the raters.

In practice we have experienced issues when fitting models via
optimisation when the non-diagonal entries of \(\beta\) are less than 1
(i.e., when \(\beta_{k', k} < 1\) for some \(k' \neq k\)). The default
hyper-parameters selected in \CRANpkg{rater} ensure that when
there are only a few possible categories---specifically, when
\(K \leqslant 4\)---then this does not occur. When \(K\) is larger than this
and the default hyper-parameters give rise to non-diagonal entries
smaller than 1, \CRANpkg{rater} will report a useful warning message.
This only arises when using optimisation; it does not arise when using MCMC.
Users who wish to use optimisation with a large number of categories can
specify different hyper-parameters to avoid the warning message.

\hypertarget{hyper-parameters-for-the-prevalence}{%
\subsubsection{Hyper-parameters for the prevalence}\label{hyper-parameters-for-the-prevalence}}

The \(\alpha\) hyper-parameters define the prior for the prevalence parameters \(\pi\).
All of these are essentially nuisance parameters and not of direct interest,
and would typically have little influence on the inference for \(\theta\) or \(z\).
The inclusion in the model of a large amount of prior information on the values
of the prevalence parameters would clearly influence the posterior distributions
of the corresponding population class frequencies. We would, however, expect
inferences about the accuracy of raters (whether this is assumed to be class-specific
or not) to depend largely on the number of ratings captured by the dataset,
not the prevalence parameters themselves. We have, therefore, not explored varying
\(\alpha\) and have simply set them to the same values as suggested in the
\emph{Stan User's Guide} (Stan Development Team 2021): \(\alpha_k = 3\) for all \(k\).

\hypertarget{hierarchical-dawidskene-model}{%
\subsection{Hierarchical Dawid--Skene model}\label{hierarchical-dawidskene-model}}

Paun et al.\ (2018) introduce a `Hierarchical Dawid--Skene' model that extends the
original one by replacing the Dirichlet prior on \(\theta_{j,k}\) (and the
associated hyper-parameter \(\beta\)) with a hierarchical specification
that allows for partial pooling, which allows the sharing of information about
raters' performance across rater-specific parameters. It requires two
\(K \times K\) matrices of parameters, \(\mu\) and \(\sigma\), with
\[
\begin{aligned}
\mu_{k, k'} &\sim \begin{cases}
    \textrm{Normal}(2, 1), & k = k' \\
    \textrm{Normal}(0, 1), & k \neq k' \\
    \end{cases}
    \quad \forall k, \ k' \in \{1, \dots, K\} \\
\sigma_{k, k'} &\sim \textrm{Half-Normal}(0, 1)
    \quad \forall k, \ k' \in \{1, \dots, K\}.
\end{aligned}
\]

We then have that:
\[
\gamma_{j, k, k'} \sim \textrm{Normal}(\mu_{k, k'}, \sigma_{k, k'})
    \quad \forall
    j \in \{1, \dots, J\},
    k, k' \in \{1, \dots K\}
\]
which are then normalised into probabilities via the multinomial logit
link function (also known as the \emph{softmax} function) in order to define
the elements of the error matrices, which are:
\[
\theta_{j, k, k'}
    = \frac{e^{\gamma_{j, k, k'}}}{\sum_{k'=1}^K e^{\gamma_{j, k, k'}}}.
\]
Other details, such as the distribution of \(z\) and \(y\), are defined in
the same way as per the Dawid--Skene model. The implementation in
\CRANpkg{rater} differs from the
implementation from Paun et al.\ (2018) by modifying the prior parameters to
encode the assumption that the raters are generally accurate.
Specifically, the higher means of the diagonal elements of \(\mu\) encode
that the raters are accurate, as after transformation by the softmax
function larger values of \(\mu\) will produce higher probabilities in
\(\theta\). The assumption that the raters are accurate is also encoded in
the prior distribution for \(\theta\) in the Dawid--Skene model described
above.

One interpretation of the hierarchical model is as a `partial pooling'
compromise between the full Dawid--Skene model and the model with only
one rater (see \protect\hyperlink{sec:homogenous}{Section 4.5}). This is depicted in
\autoref{fig:relationships}. Another interpretation of the model is that it
treats the accuracy of raters given a specific latent class as a random
effect, not a fixed effect as in the Dawid--Skene model. Paun et al.\ (2018) show,
via a series of simulations and examples, that this hierarchical model
generally improves the quality of predictions over the original
Dawid--Skene model.

\hypertarget{class-conditional-dawidskene}{%
\subsection{Class-conditional Dawid--Skene}\label{class-conditional-dawidskene}}

Another variant of the Dawid--Skene model imposes constraints on the
entries in the error matrices of the raters. Given the presumption of
accurate raters, one natural way to do this is to force all non-diagonal
entries in a given row of the error matrix to take a constant value that
is smaller than the corresponding diagonal entry. Formally the model is
the same as the Dawid--Skene model except that we have:
\[
    \theta_{j, k, k'} =
        \begin{cases}
            p_{j, k} & \text{if}\ k = k' \\
            \dfrac{1 - p_{j,k}}{(K - 1)} & \text{otherwise.}
        \end{cases}
\]
To make the model fully Bayesian we use the following prior probability
distributions:
\[
    p_{j, k} \sim \textrm{Beta}(\beta_{1,k}, \beta_{2,k}),
        \quad \forall j \in {1, \dots, J}.
\]

In \CRANpkg{rater} we set
\(\beta_{1, k} = N p\) and \(\beta_{2, k} = N(1 - p)\) for all \(k\), where
\(N = 8\) and \(p = 0.6\) as in \protect\hyperlink{sec:dawid-skene}{Section 4.1}. While the
constraints in this model may lead to imprecise or biased estimation of
parameters for any raters that are substantially inaccurate or even
antagonistic, the reduced number of parameters of this model relative to
the full Dawid--Skene model make it much easier to fit. This model was
referred to as the `class-conditional Dawid--Skene model' by Li and Yu (2014).

\hypertarget{sec:homogenous}{%
\subsection{Homogeneous Dawid--Skene model}\label{sec:homogenous}}

Another related model is the `multinomial model' described by Paun et al.\ (2018),
where the full Dawid--Skene model is constrained so that the accuracy of
all raters is assumed to be identical. The constraint can be
equivalently formulated as the assumption that all of the ratings were
done by a single rater, thus the raters and the sets of rating they
produce are exchangeable. This model can be fitted using
\CRANpkg{rater} by simply modifying the data so that only one rater rates every
item present and then fitting the usual Dawid--Skene model.

\hypertarget{relationships-between-the-models}{%
\subsection{Relationships between the models}\label{relationships-between-the-models}}

All of the models presented in this paper can be directly related to the
original Dawid--Skene model. \autoref{fig:relationships} shows the
relationships between the models implemented in the package, which are
coloured blue. The hierarchical model can be seen as a `partial pooling'
model where information about the performance of the raters is shared.
The Dawid--Skene model can then be seen as the `no pooling' extreme
where all raters' error matrices are estimated independently, while the
homogeneous model is the other extreme of complete pooling where one
error matrix is fitted for all of the raters. In addition, the
class-conditional model is a direct restriction of the Dawid--Skene
model where extra constraints are placed on the error matrices.

\hypertarget{sec:relationships-existing-models}{%
\subsection{Relationships to existing models}\label{sec:relationships-existing-models}}

The Dawid--Skene model is also closely related to existing latent class
models. Usually, in latent class models, the number of latent classes
can be chosen to maximise a suitable model selection criterion, such as
the BIC, or to make the interpretation of the latent classes easier. In
the Dawid--Skene model, however, the number of latent classes must be
fixed to the number of categories to ensure that the error matrices can
be interpreted in terms of the ability of the raters to rate the
categories that appear in the data. Some specific models to which the
Dawid--Skene model is related are depicted in \autoref{fig:relationships}
and are coloured white.

The latent class model implemented in
\CRANpkg{BayesLCA} is equivalent
to the Dawid--Skene model fitted to binary data if the number of classes
(\(G\) in the notation of White and Murphy (2014)) is set to 2. In that case, each of
the \(M\) dimensions of the binary response variable can be interpreted as
a rater, so that \(J = M\).

The Dawid--Skene model is also a special case of the model implemented
in \CRANpkg{poLCA} (Linzer and Lewis 2011). The models are equivalent if they have the
same number of latent classes (\(R = K\), where \(R\) is the number of latent
classes in the \CRANpkg{poLCA} model) and if all of
the \(J\) categorical variables in the \CRANpkg{poLCA} model take values in
the set \(\{1, \dots K\}\) (that is, \(K_j = K\) for all \(j \in J\)), so that
each of the \(J\) variables can be interpreted as the ratings of a
particular rater.

In addition, the Dawid--Skene model is a special case of the model
implemented in \CRANpkg{randomLCA} (Beath 2017). When no random effects are
specified and \(K = 2\), the model is equivalent to the Dawid--Skene model with
two classes. While \CRANpkg{randomLCA} allows
selecting an arbitrary number of latent classes, it only supports binary
data.

Finally, when \(K = 2\) the Dawid--Skene model reduces to the so-called
`traditional' or `standard' latent class model as first described by
Hui and Walter (1980). See Asselineau et al.\ (2018) for a more modern description.

\begin{figure}
\centering
\begin{tikzpicture}[node distance=2cm]
\tikzstyle{rater}=[
    rectangle,
    rounded corners,
    minimum width=2cm,
    minimum height=1cm,
    text centered,
    fill=blue!20!white,
    draw=black
]
\tikzstyle{other}=[
    rectangle,
    rounded corners,
    minimum width=2cm,
    minimum height=1cm,
    text centered,
    draw=black
]
\tikzstyle{arrow}=[
    thick,
    ->,
    >=stealth
]
\node [rater] (DS) {Dawid--Skene (DS)};
\node [rater, below of=DS] (HODS) {Homogeneous DS};
\node [rater, left=2cm of DS] (HIDS) {Hierarchical DS};
\node [other, above of=HIDS] (poLCA) {\CRANpkg{poLCA} model};
\node [other, above of=DS] (randomLCA) {\CRANpkg{randomLCA} model};
\node [rater, right=2.5cm of DS] (CCDS) {Class-conditional DS};
\node [other, above of=CCDS] (BayesLCA) {\CRANpkg{BayesLCA} model};
\node [other, below of=CCDS] (LCM) {Traditional LCM};

\draw [arrow] (DS) -- node[anchor=east] {$J = 1$} (HODS);
\draw [arrow] (randomLCA) -- node[anchor=east] {$K = 2$} (DS);
\draw [arrow] (BayesLCA.west) -- node[anchor=south, yshift = 0.2cm] {$G = 2$} (DS);
\draw [arrow] (poLCA) -- node[anchor=east, xshift = -0.4cm, yshift = -0.1cm] {$K_j = K, R = K$} (DS);
\draw [arrow] (HIDS) -- node[anchor=south] {$\sigma^2 \rightarrow \infty$} (DS);
\draw [arrow] (HIDS.south) -- node[xshift = -0.1cm, yshift = -0.1cm, anchor=east] {$\sigma^2 \rightarrow 0$} (HODS.west);
\draw [arrow] (DS) -- node[text width=2.1cm, xshift = 0.3cm, yshift = -0.2, anchor=south] {Equal off-diagonal $\theta$} (CCDS);
\draw [arrow] (DS.east) -- node[yshift = -0.5cm] {$K = 2$} (LCM.west);
\end{tikzpicture}
\caption{Relationships between models. Models coloured blue are implemented in
\CRANpkg{rater}. DS: Dawid--Skene. LCM: latent class model.}
\label{fig:relationships}
\end{figure}

\hypertarget{sec:relationships-existing-packages}{%
\subsection{Relationships to existing packages}\label{sec:relationships-existing-packages}}

These relationships to existing latent class models means the
functionality implemented in \CRANpkg{rater} is similar to that
of existing R packages for fitting these models. The features of
\CRANpkg{rater} and existing packages are summarised in \autoref{tab:package-features}.

Some further details about these relationships:

\begin{itemize}
\item
  The two Bayesian packages implement different methods of inference.
  \CRANpkg{rater} uses Hamiltonian
  Monte Carlo and optimisation of the log-posterior. \CRANpkg{BayesLCA}
  implements Gibbs sampling, the expectation--maximization (EM) algorithm and a
  Variational Bayesian approach.
\item
  Unlike the other packages, the implementation of the wide data
  format in \CRANpkg{randomLCA} supports missing data. This feature is not
  implemented in \CRANpkg{rater} because the long
  data format allows both missing data and repeated ratings.
\end{itemize}

Overall, the main features that distinguish \CRANpkg{rater} from the other
packages are its full support for the Dawid--Skene model and extensions,
and its support for repeated ratings.

\begin{table}
    \caption{Features of \CRANpkg{rater} and existing R packages for fitting
    latent class models. DS: Dawid--Skene.}
    \centering
    \begin{tabular}{lllllll}
    \toprule
    Package & DS & DS model & Fitting & Response & Repeated & Data \\
         & model & variants & method  & type     & ratings  & formats \\
    \midrule
    \addlinespace
    \CRANpkg{rater} & Yes & Yes & Bayesian & Polytomous & Yes & Wide, \\
                    &     &     &          &            &     & grouped, \\
                    &     &     &          &            &     & long \\
    \addlinespace
    \CRANpkg{BayesLCA} & When    & No & Bayesian & Binary & No & Wide, \\
                       & $K = 2$ &    &          &        &    & grouped \\
    \addlinespace
    \CRANpkg{randomLCA} & When    & No & Frequentist & Binary & No & Wide, \\
                        & $K = 2$ &    &             &        &    & grouped \\
    \addlinespace
    \CRANpkg{poLCA} & Yes & No & Frequentist & Polytomous & No & Wide \\
    \bottomrule
    \end{tabular}
    \label{tab:package-features}
\end{table}

\hypertarget{sec:implementation}{%
\section{Implementation details}\label{sec:implementation}}

\CRANpkg{rater} uses Stan (Carpenter et al.\ 2017) to fit the above models to data.
It therefore supports
both optimisation and Markov chain Monte Carlo (MCMC) sampling, in
particular using the No U-Turn Sampler (NUTS) algorithm
(Hoffman and Gelman 2014). For most datasets we recommend using NUTS, due
to the fact that it will generate realisations of the full posterior
distribution. Optimisation may be useful, however, for particularly
large datasets, especially if \(K\) is large, or if there is a need to fit
a large number of models.

\hypertarget{sec:marginalisation}{%
\subsection{Marginalisation}\label{sec:marginalisation}}

The NUTS algorithm relies on computing derivatives of the (log)
posterior distribution with respect to all parameters. It cannot,
therefore, be used when the posterior contains discrete parameters, such
as the latent class in the Dawid--Skene model. To overcome this
difficulty the Stan models implemented in
\CRANpkg{rater} use marginalised
forms of the posterior to allow NUTS to be used. In other words, for the
Dawid--Skene model we implement the likelihood
\(\Pr(y \mid \theta, \pi)\), as described earlier in
\protect\hyperlink{sec:dawid-skene}{Section 4.1}, together with priors for \(\theta\) and \(\pi\).

To avoid any confusion, we stress the fact that our choice to
marginalise over the vector \(z\) is purely because it is discrete and we
want to use the NUTS algorithm. It is not due to the fact that the
components of the vector \(z\) are inherently latent variables.
Alternative Bayesian implementations could avoid marginalisation and
sample the \({z}_{i}\)'s directly (although this may be less efficient,
see \protect\hyperlink{sec:marginalisation-effciency}{Section 6.3} for discussion). Similarly, if
the components of \(z\) were all continuous, we would be able to sample
them using NUTS. If one or more of the other non-latent parameters were
discrete, then we would need to marginalise over them too.

\hypertarget{sec:conditioning}{%
\subsection{Inference for the true class via conditioning}\label{sec:conditioning}}

Marginalising over \(z\) allows us to use NUTS and sample efficiently from
the posterior probability distribution of \(\theta\) and \(\pi\). We are
still able to obtain the posterior distribution of the components
\({z}_{i}\) of \(z\) as follows. For each posterior draw of \(\theta\) and
\(\pi\), we calculate the conditional posterior for \({z}_{i}\),
\(\Pr({z}_{i} \mid \theta^*, \pi^*, y)\), where the conditioning is on the
\textbf{drawn} values, \(\theta^*\) and \(\pi^*\). This can be done using Bayes'
theorem, for example in the Dawid--Skene model,
\[
\begin{aligned}
   \Pr(z_i = k \mid \theta^*, \pi^*, y)
&= \frac{\Pr(y \mid z_i = k, \theta^*, \pi^*)
         \Pr(z_i = k \mid \theta^*, \pi^*)}
        {\Pr(y \mid \theta^*, \pi^*)} \\
&= \frac{\pi^*_k \prod_{j = 1}^{J}
         \theta^*_{j, k, y_{i,j}}}
        {\sum_{m = 1}^{K} \pi^*_m \prod_{j = 1}^{J} \theta^*_{j, m, y_{i,j}}}.
\end{aligned}
\]
To get a Monte Carlo estimate of the marginal posterior,
\(\Pr({z}_{i} \mid y)\), we take the mean of the above conditional
probabilities across the posteriors draws of \(\theta\) and \(\pi\). To
formally justify this step we can use the same argument that justifies
the general MCMC technique of Rao--Blackwellization, see
\protect\hyperlink{sec:rao-blackwellization}{Section~6}. In practice, this calculation is
straightforward with Stan because we can write the log posterior in
terms of the (log scale) unnormalised versions of the conditional
probabilities and output them alongside the posterior draws. Then it is
simply a matter of renormalising them, followed by taking the mean
across draws.

\hypertarget{data-summarisation}{%
\subsection{Data summarisation}\label{data-summarisation}}

\CRANpkg{rater} implements the
ability to use different forms of the likelihood depending on the format
of the data that is available. For example, for a balanced design the
full likelihood for the Dawid--Skene model is
\[
\Pr(y \mid \theta, \pi)
= \prod_{i = 1}^{I}
  \left(\sum_{k = 1}^{K}
        \left(\pi_{k} \cdot \prod_{j = 1}^{J}
                            \theta_{j, k, y_{i, j}}\right)\right).
\]
Looking closely, we see that the contribution of any two items with the
same pattern of ratings, that is, the same raters giving the same
ratings, is identical. We can therefore rewrite it as a product over
patterns rather than items,
\[
\Pr(y \mid \theta, \pi)
= \prod_{l = 1}^{L}
  \left(\sum_{k = 1}^{K}
        \left(\pi_{k} \cdot \prod_{j = 1}^{J}
                            \theta_{j, k, y_{l, j}}\right)\right)^{n_l},
\]
where \(L\) is the number of distinct rating `patterns', \(n_l\) is the
number of times pattern \(l\) occurs and \(y_{l, j}\) denotes the rating
given by rater \(j\) within pattern \(l\). This corresponds directly to
having the data in the `grouped' format, see \protect\hyperlink{sec:data-formats}{Section 2}.

This rewriting is useful for balanced designs where there are many
ratings but few distinct patterns, as it removes the need to iterate
over every rating during computation of the posterior, reducing the time
for model fitting. An example of this type of data are the \texttt{caries} data
presented in \protect\hyperlink{sec:usage}{Section 7}.

This technique is not always helpful. For example, if the data has many
missing entries, the number of distinct patterns \(L\) may be similar or
equal to than the total number of ratings. Consequently, rewriting to
use the grouped format may not save much computation. For this reason,
the implementation of \CRANpkg{rater} does not allow
missing values in grouped data. Note that in the long data format we
simply drop rows that correspond to missing data, a result of which is
that missing data is never explicitly represented in \CRANpkg{rater}.

Both the \CRANpkg{randomLCA} and \CRANpkg{BayesLCA} packages support grouped
format data input. Of these, CRANpkg\{randomLCA\} supports missing values in
grouped format data.

\hypertarget{sec:rao-blackwellization}{%
\section{Marginalisation and Rao--Blackwellization}\label{sec:rao-blackwellization}}

The technique of marginalisation for sampling discrete random variables
discussed in \protect\hyperlink{sec:implementation}{Section 5} is well known in the Stan
community. We feel, however, that the interpretation and use of the
relevant conditional probabilities and expectations has not been clearly
described before, so we attempt to do so here. We begin with a brief
review of the theory of marginalisation, which is a special case of a
more general technique often referred to as `Rao--Blackwellization'. See
Owen (2013) for an introduction and Robert and Roberts (2021) for a
recent survey of the use of Rao--Blackwellization in MCMC.

\hypertarget{sec:rao-blackwell-connection}{%
\subsection{Connection with the Rao--Blackwell theorem}\label{sec:rao-blackwell-connection}}

Suppose we are interested in estimating \(\mu = \mathbb{E}(f(X, Y))\) for
some function \(f\) of random variables \(X\) and \(Y\). An example of such an
expectation is \(\mu = \mathbb{E}(Y)\) where we take \(f(x, y) = y\). If we
can sample from the joint distribution of \(X\) and \(Y\), \(p_{X, Y}(x, y)\),
the obvious Monte Carlo estimator of this expectation is
\[
\hat{\mu} = \frac{1}{n} \sum_{i = 1}^{n} f(x_i, y_i)
\]
where \((x_i, y_i)\) are samples from the joint distribution. Now let
\(h(x) = \mathbb{E}(f(X, Y) \mid X = x)\). An alternate conditional
estimator---the so-called \emph{Rao--Blackwellized estimator}---of \(\mu\) is
\[
\hat{\mu}_{\textrm{cond}} = \frac{1}{n} \sum_{i = 1}^{n}
h(x_i)
\]
where the \(x_i\) are sampled from the distribution of \(X\). This estimator
is justified by the fact that,
\[
\mathbb{E}(h(X)) = \mathbb{E}(\mathbb{E}(f(X, Y) \mid X)) = \mathbb{E}(f(X,Y)) = \mu
\]
For this to be effective we need to be able to efficiently compute
(ideally, analytically) the conditional expectation \(h(X)\). The original
motivation for using this technique was to reduce the variance of Monte
Carlo estimators. This follows from noting that
\[
\begin{aligned}
    \mathop{\mathrm{var}}(f(X, Y))
    &= \mathbb{E}(\mathop{\mathrm{var}}(f(X, Y) \mid X)) + \mathop{\mathrm{var}}(\mathbb{E}(f(X, Y) \mid X)) \\
    &= \mathbb{E}(\mathop{\mathrm{var}}(f(X, Y) \mid X)) + \mathop{\mathrm{var}}(h(X)),
\end{aligned}
\]
and since the first term on the right-hand side is non-negative, we have
\(\texttt{(}f(X, Y)) \geqslant \texttt{(}h(X))\).

From this result it follows that, if the draws \(x_i\) are independent
samples, then
\[
\mathop{\mathrm{var}}(\hat{\mu}_{\textrm{cond}}) < \mathop{\mathrm{var}}(\hat{\mu})
\]
for all functions \(f(X, Y)\). In the applications we consider here,
however, \(x_i\) will be drawn using MCMC and therefore draws will not be
independent. In this case it is possible to construct models and
functions \(f(\cdot)\) for which conditioning can increase the variance of
the estimator (Geyer 1995).

The name `Rao--Blackwellization' arose due to the similarity of the
above argument to that used in the proof of the Rao--Blackwell theorem
(Blackwell 1947), which states that given an estimator \(\hat{\mu}\) of a
parameter \(\mu\), the conditional expectation of \(\hat{\mu}\) given a
sufficient statistic for \(\mu\) is potentially more efficient and
certainly no less efficient than \(\hat{\mu}\) as an estimator of \(\mu\).
The Rao--Blackwellized estimators presented above do not, however, make
any use of sufficiency and do not have the same optimality guarantees
that the Rao--Blackwell theorem provides, making the name less than apt.
Following Geyer (1995), we prefer to think of the estimators presented
above as simply being formed from averaging conditional expectations.

\hypertarget{marginalisation-in-stan}{%
\subsection{Marginalisation in Stan}\label{marginalisation-in-stan}}

The motivation for marginalisation in Stan is to enable estimation of
\(\mu = \mathbb{E}(f(X, Y))\) without having to sample from \((X, Y)\) if
either \(X\) or \(Y\) is discrete. Suppose that \(Y\) is discrete and \(X\)
continuous. To compute a Monte Carlo estimate of \(\mathbb{E}(f(X, Y))\)
using Stan we carry out four steps.

First, marginalise out \(Y\) from \(p_{X, Y}(x, y)\) to give \(p_{X}(x)\).
(See Equation \eqref{eq:eqnmarginalise} in \protect\hyperlink{sec:dawid-skene}{Section 4.1} for how this is
done in the Dawid--Skene model.) This marginal distribution, which only
contains continuous parameters, should then be implemented as a Stan
model. Second, using Stan as usual, sample from the distribution of the
continuous parameters \(X\) to give Monte Carlo samples \(\{x_i\}\). Given
only the samples from the distribution of \(X\), we can estimate
\(\mathbb{E}(f(X, Y))\) using the Rao--Blackwellized estimator described
in the previous section. Doing so requires us to evaluate
\(h(X) = \mathbb{E}(f(X, Y) \mid X)\) for all functions \(f(\cdot)\) in
which we may be interested; this is the third step of the process.
Finally, as the fourth step, we can evaluate \(h(x)\) for each of the
Monte Carlo draws \(x_i\) and estimate \(\mathbb{E}(f(X, Y))\) by
\[
\frac{1}{n} \sum_{i = 1}^n h(x_i).
\]

\hypertarget{evaluating-the-conditional-expectation}{%
\subsubsection{Evaluating the conditional expectation}\label{evaluating-the-conditional-expectation}}

The third step outlined above may appear to require substantial
mathematical manipulation. In practice, however, we can use the discrete
nature of the latent class to simplify the calculation. Specifically,
for any function \(f(x, y)\) we have
\[
h(X)
    = \mathbb{E}(f(X, Y) \mid X)
    = \sum_{k} f(X, k) \Pr(Y = k \mid X),
\]
where we sum over the values in the latent discrete parameters. An
essential part of this formula is the probability of the discrete
parameters conditional on the continuous parameters,
\(\Pr(Y = k \mid X)\). This quantity can be derived easily through Bayes'
theorem or can be encoded as part of the marginalised Stan model; see
\protect\hyperlink{sec:conditioning}{Section 5.2} or the next section for how this is done in the
case of the Dawid--Skene model.

In the Dawid--Skene model, and many other models with discrete
variables, the discrete variables only take values in a finite set. Some
models however may have discrete parameters which take countably
infinite values, such as Poisson distributions. In this case,
marginalisation is still possible but will generally involve
approximating infinite sums, both when marginalising out the discrete
random variables and calculating the expectations (as in the equation
above), which may be computationally expensive.

Furthermore, for the cases where \(f(x, y)\) is a function of only \(x\) or
\(y\), the general formula simplifies further. Firstly, when
\(f(x, y) = f(x)\) we have
\[
h(X)
    = \sum_{k} f(X) \Pr(Y = k \mid X)
    = f(X) \sum_{k} \Pr(Y = k \mid X)
    = f(X).
\]
This means that we can estimate \(\mathbb{E}(f(X))\) with the standard,
seemingly unconditional, estimator:
\[
\sum_{i = 1}^n f(x_i).
\]
Even after marginalisation, computing expectations of functions of the
continuous parameters can be performed as if no marginalisation had
taken place.

Secondly, when \(f(x, y) = f(y)\) we have that
\[
h(X) = \sum_{k} f(k) \Pr(Y = k \mid X).
\]

An important special case of this result is when
\(f(x, y) = \mathop{\mathrm{\mathbf{1}}}(y = k)\), where
\(\mathop{\mathrm{\mathbf{1}}}\) is the indicator function. This is
important because it allows us to recover the probability mass function
of the discrete random variable \(Y\), since
\(\mathbb{E}(f(X, Y)) = \mathbb{E}(\mathop{\mathrm{\mathbf{1}}}(Y = k)) = \Pr(Y = k)\).
In this case we have
\[
h(X)
    = \sum_{k} \mathop{\mathrm{\mathbf{1}}}(y = k) \Pr(Y = k \mid X)
    = \Pr(Y = k \mid X).
\]
We therefore estimate \(\Pr(Y = k)\) with:
\[
\frac{1}{n} \sum_{i = 1}^n
    \Pr(Y = k \mid X = x_i).
\]
Again, we stress that our ability to do these calculations relies upon
being able to easily compute \(\Pr(Y = k \mid X = x_i)\) for each of the
Monte Carlo draws \(x_i\).

\hypertarget{estimating-the-conditional-probability-of-the-true-class}{%
\subsubsection{Estimating the conditional probability of the true class}\label{estimating-the-conditional-probability-of-the-true-class}}

For the categorical rating problem (using the Dawid--Skene model), the
discrete random variable of interest for each item \(i\) is the true class
of the item, \({z}_{i}\). We use the technique from the previous section
to calculate the posterior probability of the true class, as described
in \protect\hyperlink{sec:conditioning}{Section 5.2}. In this case, the discrete variable is
\(z_i\) (taking on the role that \(Y\) played in the previous section), the
continuous variables are \(\theta\) and \(\pi\) (which together take on the
role that \(X\) played in the previous section), and all probability
calculations are always conditional on the data (the ratings, \(y\)).

\hypertarget{sec:marginalisation-effciency}{%
\subsection{Efficiency of marginalisation}\label{sec:marginalisation-effciency}}

It is not immediately obvious whether marginalisation is more or less
efficient than techniques that actually realise a value for the discrete
random variable at each iteration, such as Gibbs sampling. Marginalising
can be viewed as a form of Rao--Blackwellization, a general MCMC
technique designed to reduce the variability of Monte Carlo estimators.
This link strongly suggests that marginalisation is more efficient than
using discrete random variables. Unfortunately, limitations of the
theory of Rao--Blackwellization (see
\protect\hyperlink{sec:rao-blackwell-connection}{Section 6.1}) and the difficulty of
theoretically comparing different sampling algorithms, such as Gibbs
sampling and NUTS, means that it is unclear whether marginalisation will
always be computationally superior for a given problem.

However, in practice marginalisation does seem to improve convergence at
least for non-conjugate models. For example, Yackulic et al.\ (2020) show that for
the Cormack--Jolly--Seber model marginalisation greatly speeds up
inference in JAGS, BUGS and WinBUGS. They also demonstrate that the
marginalised model implemented in Stan is orders of magnitude more
efficient than the marginalised models in the other languages. These
results show that marginalisation has the potential to speed up classic
algorithms and also allows the use of more efficient gradient-based
algorithms, such as NUTS, for problems involving discrete random
variables.

A recent similar exploration of marginalisation for some simple mixture
models and the Dawid--Skene model, using JAGS and Stan, suggests that
the software implementation has a greater impact on efficiency than the
choice of whether or not marginalisation is used (Zhang et al.\ 2022). In their
comparisons, Stan usually achieved the best performance (and note that
Stan requires marginalisation of discrete parameters).

For these reasons, we recommend that practitioners using models with
discrete parameters consider marginalisation if more efficiency is
desired, and implement their models using Stan rather than JAGS.

\hypertarget{sec:usage}{%
\section{Example usage}\label{sec:usage}}

To demonstrate \CRANpkg{rater} we use two example datasets.

The first dataset is taken from the original paper introducing the Dawid--Skene
model (Dawid and Skene 1979). The data consist of ratings, on a 4-point scale, made
by five anaesthetists of patients' pre-operative health. The ratings were based
on the anaesthetists assessments of a standard form completed for all of the
patients. There are 45 patients (items) and four anaesthetists (raters) in
total. The first anaesthetist assessed the forms a total of three times, spaced
several weeks apart. The other anaesthetists each assessed the forms once. As
in the Dawid--Skene paper, we will not seek to model the effect of time on the
ratings of the first anaesthetist.

First we load the \CRANpkg{rater} package:

\begin{verbatim}
library(rater)
\end{verbatim}

\begin{verbatim}
#> * The rater package uses `Stan` to fit bayesian models.
#> * If you are working on a local, multicore CPU with excess RAM please call:
#> * options(mc.cores = parallel::detectCores())
#> * This will allow Stan to run inference on multiple cores in parallel.
\end{verbatim}

This will display information about altering options used by Stan to
make best use of computational resources on your machine. We can then load and
look at a snippet of the anaesthesia data, which is included in the package.

\begin{verbatim}
data("anesthesia", package = "rater")
head(anesthesia)
\end{verbatim}

\begin{verbatim}
#>   item rater rating
#> 1    1     1      1
#> 2    1     1      1
#> 3    1     1      1
#> 4    1     2      1
#> 5    1     3      1
#> 6    1     4      1
\end{verbatim}

These data are arranged in `long' format where each row corresponds to a single
rating. The first column gives the index of the item that was rated, the second
the index of the rater that rated the item and the third the actual rating that
was given. This layout of the data supports raters rating the same item
multiple times, as happens in the dataset. It is also the most convenient from
the perspective of fitting models but may not always be the optimal way to
store or represent categorical rating data; see
\protect\hyperlink{sec:using-grouped-data}{Section 7.6} for an example using the `grouped' data
format.

\hypertarget{fitting-the-model}{%
\subsection{Fitting the model}\label{fitting-the-model}}

We can fit the Dawid--Skene model using MCMC by running the command:

\begin{verbatim}
fit_1 <- rater(anesthesia, dawid_skene())
\end{verbatim}

This command will print the running output from Stan, providing an
indication of the progress of the sampler (for brevity, we have not shown this
output here).
To fit the model via optimisation, we set the \texttt{method} argument to
\texttt{"optim"}:

\begin{verbatim}
fit_2 <- rater(anesthesia, dawid_skene(), method = "optim")
\end{verbatim}

The second argument of the \texttt{rater()} function specifies the model to use.
We have implemented this similarly to the \texttt{family} argument in
\texttt{glm()}, which can be passed as either a function or as a character
string. The above examples pass the model as a function. We could have instead
passed it as a string, for example:

\begin{verbatim}
fit_2 <- rater(anesthesia, "dawid_skene", method = "optim")
\end{verbatim}

Either version will fit the Dawid--Skene model using the default choice of
prior. The benefit of passing the model as a function is that it allows you to
change the prior, see \protect\hyperlink{sec:different-models-priors}{Section 7.5}.

\hypertarget{inspecting-the-fitted-model}{%
\subsection{Inspecting the fitted model}\label{inspecting-the-fitted-model}}

\CRANpkg{rater} includes several ways to inspect the output of fitted models.
These are summarised in \autoref{tab:inspecting-fitted-models}, and we
illustrate many of them here. Firstly, we can generate a text summary:

\begin{verbatim}

summary(fit_1)
\end{verbatim}

\begin{verbatim}
#> Model:
#> Bayesian Dawid and Skene Model 
#> 
#> Prior parameters:
#> 
#> alpha: default
#> beta: default
#> 
#> Fitting method: MCMC
#> 
#> pi/theta samples:
#>                mean   5%  95% Rhat ess_bulk
#> pi[1]          0.37 0.27 0.48    1  9094.91
#> pi[2]          0.41 0.30 0.52    1  8050.47
#> pi[3]          0.14 0.07 0.23    1  6864.65
#> pi[4]          0.07 0.03 0.14    1  6665.16
#> theta[1, 1, 1] 0.86 0.79 0.93    1  7728.01
#> theta[1, 1, 2] 0.10 0.05 0.17    1  7857.38
#> theta[1, 1, 3] 0.02 0.00 0.05    1  5619.81
#> theta[1, 1, 4] 0.02 0.00 0.05    1  6283.33
#> # ... with 76 more rows
#> 
#> z:
#>      MAP Pr(z = 1) Pr(z = 2) Pr(z = 3) Pr(z = 4)
#> z[1]   1      1.00      0.00      0.00      0.00
#> z[2]   3      0.00      0.00      0.98      0.02
#> z[3]   2      0.38      0.62      0.00      0.00
#> z[4]   2      0.01      0.99      0.00      0.00
#> z[5]   2      0.00      1.00      0.00      0.00
#> z[6]   2      0.00      1.00      0.00      0.00
#> z[7]   1      1.00      0.00      0.00      0.00
#> z[8]   3      0.00      0.00      1.00      0.00
#> # ... with 37 more items
\end{verbatim}

This function will show information about which model has been fitted and the
values of the prior parameters that were used. The displayed text also contains
information about the parameter estimates and posterior distributions, and the
convergence of the sampler.

\begin{table}
    \centering
    \caption{Methods to inspect fitted models and what values they return.}
    \begin{tabular}{lll}
    \toprule
    \textbf{Function}
        &  \textbf{MCMC mode}   &  \textbf{Optimisation mode}  \\
    \midrule
    \code{summary()}
        &  Basic information about
                                &  Basic information about  \\
        &  the fitted model
                                &  the fitted model  \\
    \addlinespace
    \code{point\_estimate()}
        &  Posterior means for $\pi$ and $\theta$,
                                &  Posterior modes for  \\
        &  posterior modes for $z$
                                &  all quantities  \\
    \addlinespace
    \code{posterior\_interval()}
        &  Credible intervals for $\pi$ and $\theta$
                                &  N/A  \\
    \addlinespace
    \code{posterior\_samples()}
        &  MCMC draws for $\pi$ and $\theta$
                                &  N/A  \\
    \addlinespace
    \code{mcmc\_diagnostics()}
        &  MCMC convergence diagnostics
                                &  N/A  \\
        &  for $\pi$ and $\theta$  \\
    \addlinespace
    \code{class\_probabilities()}
        &  Posterior distribution for $z$
                                &  Posterior distribution for $z$  \\
        &                       &  conditional on the posterior    \\
        &                       &  modes for $\pi$ and $\theta$    \\
    \bottomrule
    \end{tabular}
    \label{tab:inspecting-fitted-models}
\end{table}

We can extract point estimates for any of the parameters or the latent classes
via the \texttt{point\_estimate()} function. For example, the following will
return the latent class with the highest posterior probability (i.e., the
posterior modes) for each item:

\begin{verbatim}
point_estimate(fit_1, "z")
\end{verbatim}

\begin{verbatim}
#> $z
#>  1  2  3  4  5  6  7  8  9 10 11 12 13 14 15 16 17 18 19 20 21 22 23 24 25 26 
#>  1  3  2  2  2  2  1  3  2  2  4  2  1  2  1  1  1  1  2  2  2  2  2  2  1  1 
#> 27 28 29 30 31 32 33 34 35 36 37 38 39 40 41 42 43 44 45 
#>  2  1  1  1  1  3  1  2  2  3  2  2  3  1  1  1  2  1  2
\end{verbatim}

\pagebreak

The following will return the posterior means of the prevalence probabilities:

\begin{verbatim}
point_estimate(fit_1, "pi")
\end{verbatim}

\begin{verbatim}
#> $pi
#> [1] 0.37432122 0.40674116 0.14406295 0.07487466
\end{verbatim}

From the outputs above, we can see that the model has inferred that most
patients have health that should be classified as category 1 or 2 (roughly 40\%
each), and categories 3 and 4 being rarer (about 14\% and 7\% respectively).
Based on the point estimates, there are examples of patients from each category
in the dataset.

The function call \texttt{point\_estimate(fit1,\ "theta")} will return posterior
means for the parameters in the error matrices (not shown here, for brevity).
When used with optimisation fits, \texttt{point\_estimate()} will return
posterior modes rather than posterior means for the parameters.

\hypertarget{inspecting-posterior-distributions}{%
\subsection{Inspecting posterior distributions}\label{inspecting-posterior-distributions}}

To represent uncertainty, we need to look beyond point estimates. The function
\texttt{posterior\_interval()} will return credible intervals for the parameters.
For example, 80\% credible intervals for the elements of the error matrices:

\begin{verbatim}
head(posterior_interval(fit_1, 0.8, "theta"))
\end{verbatim}

\begin{verbatim}
#>                        10%        90%
#> theta[1, 1, 1] 0.805074415 0.91666962
#> theta[1, 1, 2] 0.055802572 0.15309851
#> theta[1, 1, 3] 0.001813371 0.03941595
#> theta[1, 1, 4] 0.001760845 0.03832660
#> theta[1, 2, 1] 0.026218408 0.10590753
#> theta[1, 2, 2] 0.791396057 0.90637035
\end{verbatim}

The function \texttt{posterior\_samples()} will return the actual MCMC draws. For
example (here just illustrating the draws of the \(\pi\) parameter):

\begin{verbatim}
head(posterior_samples(fit_1, "pi")$pi)
\end{verbatim}

\begin{verbatim}
#>           
#> iterations      [,1]      [,2]      [,3]       [,4]
#>       [1,] 0.4830595 0.3579169 0.1254949 0.03352871
#>       [2,] 0.2939839 0.3834238 0.2516929 0.07089945
#>       [3,] 0.3636097 0.4148082 0.1309119 0.09067012
#>       [4,] 0.3798456 0.3133947 0.2353036 0.07145615
#>       [5,] 0.3639428 0.4291340 0.1760525 0.03087066
#>       [6,] 0.3845465 0.3831443 0.1721656 0.06014354
\end{verbatim}

Neither of the above will work for optimisation fits, which are limited to
point estimates only. For the latent classes, we can produce the probability
distribution as follows:

\begin{verbatim}
head(class_probabilities(fit_1))
\end{verbatim}

\begin{verbatim}
#>    
#>             [,1]         [,2]         [,3]         [,4]
#>   1 9.999994e-01 1.340790e-07 2.916232e-08 4.521956e-07
#>   2 8.469190e-08 2.570243e-05 9.773942e-01 2.257999e-02
#>   3 3.823085e-01 6.171346e-01 1.075246e-04 4.493620e-04
#>   4 5.252130e-03 9.942742e-01 3.418545e-04 1.318565e-04
#>   5 2.473083e-07 9.999633e-01 3.338995e-05 3.062785e-06
#>   6 1.674456e-06 9.993819e-01 5.813534e-04 3.502596e-05
\end{verbatim}

This works for both MCMC and optimisation fits. For the former, the output is
the posterior distribution on \(z\), while for the latter it is the distribution
conditional on the point estimates of the parameters (\(\pi\) and \(\theta\)).

\hypertarget{plots}{%
\subsection{Plots}\label{plots}}

It is often easier to interpret model outputs visually.
Using \CRANpkg{rater}, we can plot the parameter estimates and
distribution of latent classes.

The following command will visualise the posterior means of the error rate
parameters, with the output shown in \autoref{fig:plot-theta}:

\begin{verbatim}
plot(fit_1, "raters")
\end{verbatim}

\begin{figure}

{\centering \includegraphics{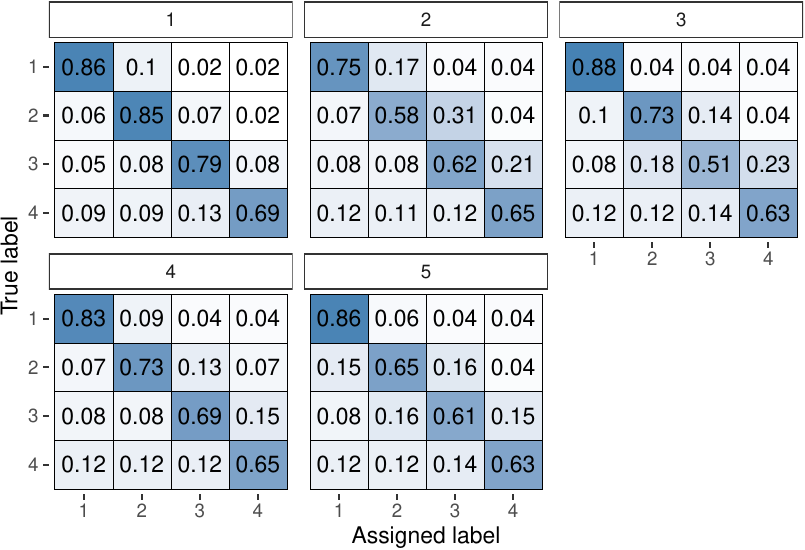} 

}

\caption{Visual representation of the inferred parameters in the error matrices ($\theta$) for the Dawid--Skene model fitted via MCMC to the anaesthesia dataset. The values shown are posterior means, with each cell shaded on a gradient from white (value close to 0) to dark blue (value close to 1).}\label{fig:plot-theta}
\end{figure}

We can see high values along the diagonals of these matrices, which indicates
that each of the 5 anaesthetists is inferred as being fairly accurate at rating
pre-operative health. Looking at the non-diagonal entries, we can see that
typically the most common errors are 1-point differences on the rating scale.

The following command visualises the latent class probabilities, with the
output shown in \autoref{fig:plot-z}:

\begin{verbatim}
plot(fit_1, "latent_class", item_index = c(2, 3, 12, 36, 38))
\end{verbatim}

\begin{figure}

{\centering \includegraphics{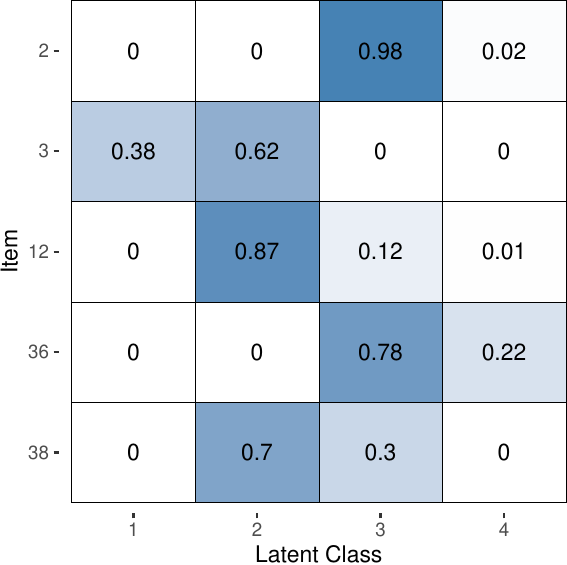} 

}

\caption{Visualisation of the inferred probability of each latent class, for a selected subset of items, for the Dawid--Skene model fitted via MCMC to the anaesthesia dataset.}\label{fig:plot-z}
\end{figure}

For the purpose of illustration, for this plot we have selected the 5 patients
with the greatest uncertainty in their pre-operative health (latent class). The
other 40 patients all have almost no posterior uncertainty for their
pre-operative health. Thus, suppose we wished to use the model to infer the
pre-operative health by combining all of the anaesthetists' ratings, we can do
so confidently for all but a handful of patients.

\begin{figure}

{\centering \includegraphics{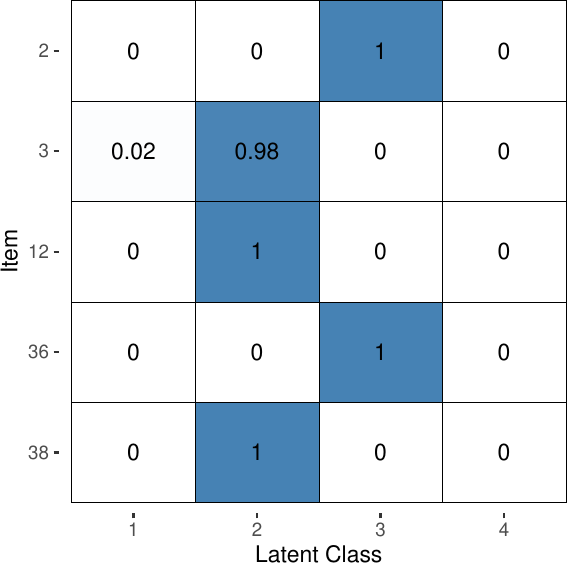} 

}

\caption{Visualisation of the inferred probability of each latent class, for a selected subset of items, for the Dawid--Skene model fitted via optimisation to the anaesthesia dataset.  Compare with \autoref{fig:plot-z}, which used MCMC instead of optimisation.}\label{fig:plot-z-fit2}
\end{figure}

The same output for the models fitted via
optimisation rather than MCMC (using \texttt{fit\_2} instead of \texttt{fit\_1}) are
shown in \autoref{fig:plot-theta-fit2} and \autoref{fig:plot-z-fit2}. We can
see that this estimation method leads to the same broad conclusions, however the
optimisation-based estimates have considerably less uncertainty: they are
typically closer to 0 or 1. This behaviour reflects the fact that
optimisation-based inference will usually not capture the full uncertainty and
will lead to overconfident estimates. Thus, we recommend using MCMC (which we
have set as the default).

\begin{figure}

{\centering \includegraphics{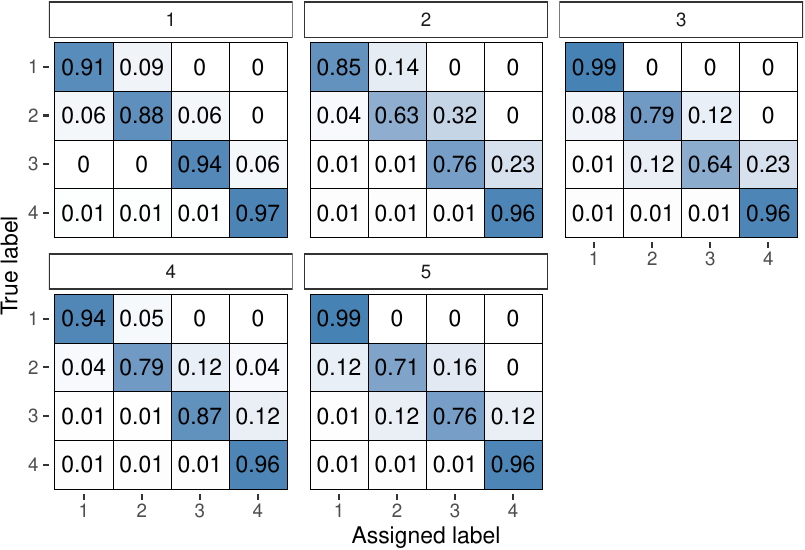} 

}

\caption{Visual representation of the inferred parameters in the error matrices ($\theta$) for the Dawid--Skene model fitted via optimisation to the anaesthesia dataset.  Compare with \autoref{fig:plot-theta}, which used MCMC instead of optimisation.}\label{fig:plot-theta-fit2}
\end{figure}

While it is typically of less interest, it is also possible to visualise the
prevalence estimates, together with credible intervals, using the following
command (see output in \autoref{fig:plot-pi}):

\begin{verbatim}
plot(fit_1, "prevalence")
\end{verbatim}

\begin{figure}

{\centering \includegraphics{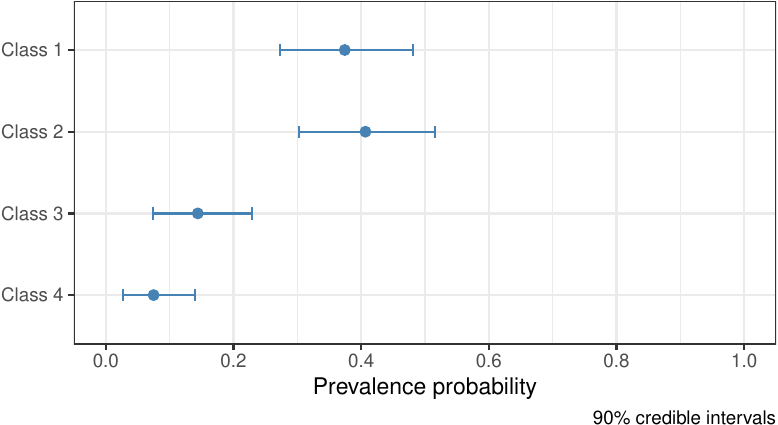} 

}

\caption{Visualisation of the inferred population prevalence parameters, with 90\% credible intervals, for the Dawid--Skene model fitted to the anaesthesia dataset.}\label{fig:plot-pi}
\end{figure}

\hypertarget{sec:different-models-priors}{%
\subsection{Different models and priors}\label{sec:different-models-priors}}

The second argument to the \texttt{rater()} function specifies what model to use,
including details of the prior if not using the default one. This gives a
unified place to specify both the model and prior.

For example, this is how to set a different prior using the Dawid--Skene model:

\begin{verbatim}
diff_alpha_fit <- rater(anesthesia, dawid_skene(alpha = rep(10, 4)))
\end{verbatim}

When specifying the \(\beta\) hyper-parameters for the Dawid--Skene model, the
user can either specify a single matrix or a 3-dimensional array. If only a matrix
specified, it will be interpreted as the hyper-parameter values for all raters.
This is useful in the common situation where the overall quality of the raters
is known but there is no information on the quality of any specific rater.
When a 3-dimensional array is passed, it is taken to specify the hyper-parameters
for each of the raters (i.e., a separate matrix for each rater). This is
useful when prior information about the quality of specific raters is
available, for example when some `raters' are diagnostic tests with known
performance characteristics.

This is how to use the class-conditional Dawid--Skene model (with default prior):

\begin{verbatim}
diff_model_fit <- rater(anesthesia, class_conditional_dawid_skene())
\end{verbatim}

Compared with the Dawid--Skene model, the latter uses error matrices with the
constraint that all off-diagonal entries must be the same. We can visualise
the \(\theta\) parameter using the following command (output in
\autoref{fig:plot-theta-diff-model}) and see that all off-diagonal elements are
indeed equal:

\begin{verbatim}
plot(diff_model_fit, "theta")
\end{verbatim}

\begin{figure}

{\centering \includegraphics{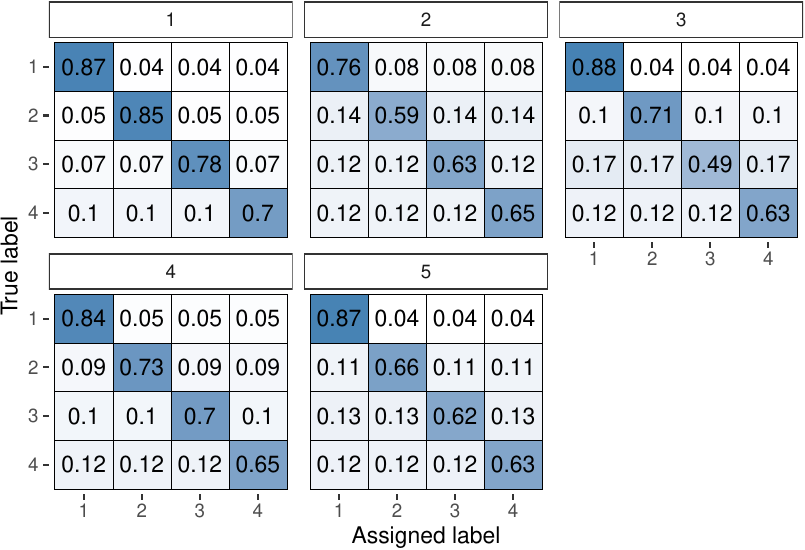} 

}

\caption{Visual representation of the inferred parameters in the error matrices ($\theta$) for the class-conditional Dawid--Skene model fitted via MCMC to the anaesthesia dataset.  Compare with \autoref{fig:plot-theta}, which used the standard Dawid--Skene model.}\label{fig:plot-theta-diff-model}
\end{figure}

\hypertarget{sec:using-grouped-data}{%
\subsection{Using grouped data}\label{sec:using-grouped-data}}

The second example dataset is taken from Espeland and Handelman (1989). It consists of
3,689 binary ratings, made by 5 dentists, of whether a given tooth was healthy
or had caries/cavities. The ratings were performed using X-ray
only, which was thought to be more error-prone than visual/tactile assessment
of each tooth (see Handelman et al.\ (1986) for more information and a description
of the wider dataset from which these binary ratings were taken).

\begin{verbatim}
data("caries", package = "rater")
head(caries)
\end{verbatim}

\begin{verbatim}
#>   rater_1 rater_2 rater_3 rater_4 rater_5    n
#> 1       1       1       1       1       1 1880
#> 2       1       1       1       1       2  789
#> 3       1       1       1       2       1   43
#> 4       1       1       1       2       2   75
#> 5       1       1       2       1       1   23
#> 6       1       1       2       1       2   63
\end{verbatim}

This is an example of `grouped' data. Each row represents a particular ratings
`pattern', with the final column being a tally that records how many instances
of that pattern appear in the data. \CRANpkg{rater} accepts data in either
`long', `wide' or `grouped' format. The `long' format is the default because
it can represent data with repeated ratings. When available the `grouped'
format can greatly speed up the computation of certain models and is convenient
for datasets that are already
recorded in that format. The `wide' format doesn't provide any computational
advantages but is a common format for data without repeated ratings and so is
provided for convenience.

Here's how we fit the model to grouped data (note the \texttt{data\_format}
argument):

\begin{verbatim}
fit3 <- rater(caries, dawid_skene(), data_format = "grouped")
\end{verbatim}

From there, we can inspect the model fit in the same way as before. For
example (output shown in \autoref{fig:plot-theta-caries}):

\begin{verbatim}
plot(fit3, "raters")
\end{verbatim}

\begin{figure}

{\centering \includegraphics{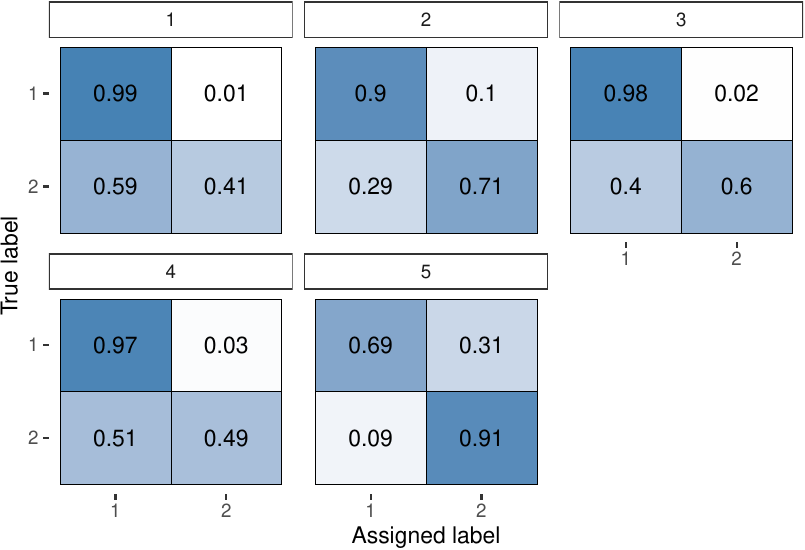} 

}

\caption{Visual representation of the inferred parameters in the error matrices ($\theta$) for the Dawid--Skene model fitted via MCMC to the caries dataset.}\label{fig:plot-theta-caries}
\end{figure}

The first 4 dentists are highly accurate at diagnosing healthy teeth (rating \(= 1\)), but less so at diagnosing caries (rating \(= 2\)). In contrast, the 5th
dentist was much better at diagnosing caries than healthy teeth.

\hypertarget{sec:convergence-diagnostics}{%
\subsection{Convergence diagnostics}\label{sec:convergence-diagnostics}}

A key part of applied Bayesian statistics is assessing whether the
MCMC sampler has converged on the posterior distribution. To summarise the
convergence of a model fit using MCMC, \CRANpkg{rater} provides the
\texttt{mcmc\_diagnostics()} function.

\begin{verbatim}
head(mcmc_diagnostics(fit_1))
\end{verbatim}

\begin{verbatim}
#>                     Rhat ess_bulk
#> pi[1]          0.9997012 9094.906
#> pi[2]          1.0009095 8050.465
#> pi[3]          1.0000406 6864.648
#> pi[4]          1.0021824 6665.159
#> theta[1, 1, 1] 1.0013982 7728.007
#> theta[1, 1, 2] 1.0011678 7857.378
\end{verbatim}

This function calculates and displays the \(\hat{R}\) statistic (\texttt{Rhat}) and bulk
effective sample size (\texttt{ess\_bulk}) for all of the \(\pi\) and \(\theta\)
parameters. Users can then check for the convergence of specific parameters by
applying standard rules, such as considering that convergence has been reached
for a specific parameter if it's \(\hat{R} < 1.01\) (Vehtari et al.\ 2021).

Users wishing to calculate other metrics, or produce convergence visualisations
such as trace plots, can either extract the underlying Stan model from the
\CRANpkg{rater} fit using the \texttt{get\_stanmodel()} function, or convert the
fit into a \texttt{mcmc.list} from the \CRANpkg{coda} package using the
\texttt{as\_mcmc.list()} function. Functions in the \CRANpkg{rstan} and
\CRANpkg{coda} packages will then allow visualisation based assessment of
convergence and the calculation of other, more advanced, diagnostics.

\begin{figure}

{\centering \includegraphics{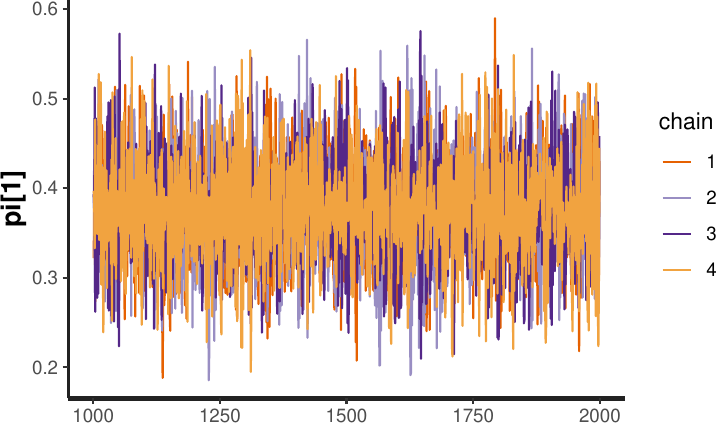} 

}

\caption{A trace plot for the $\pi_1$ parameter from the Dawid--Skene model fitted via MCMC to the anaesthesia dataset.}\label{fig:plot-rstan-traceplot}
\end{figure}

For example, the following code uses \CRANpkg{rstan} to draw the trace plot
(shown in \autoref{fig:plot-rstan-traceplot}) for one of the parameters from
the Dawid--Skene model fitted to the anaesthesia data. We can see that the
four chains show good mixing. The trace plots for the other parameters
(not shown here) are similar, indicating that the MCMC sampling procedure has
converged.

\begin{verbatim}
stan_fit <- get_stanfit(fit_1)
rstan::traceplot(stan_fit, pars = "pi[1]")
\end{verbatim}

\hypertarget{sec:model-assessment-comparison}{%
\subsection{Model assessment and comparison}\label{sec:model-assessment-comparison}}

Finally, \CRANpkg{rater} provides facilities to assess and compare fitted
models. First, it provides the \texttt{posterior\_predict()} function to simulate from
the posterior predictive distributions of all the models implemented in
\CRANpkg{rater}. The simulated data can then be used, for example, to perform
posterior predictive checks. These checks compare the data
simulated from the fitted model to the observed data. A number of datasets
are simulated from the posterior predictive distribution and a summary statistic
is calculated for each dataset. The same summary statistic is calculated on the
observed dataset and is compared to the distribution of simulated statistics.
A large discrepancy between the simulated statistics and the observed statistic
indicates possible model misspecification.

To illustrate this functionality, we consider assessing the fit of the
Dawid--Skene model to the anaesthesia dataset. In this example, we simulate
1,000 datasets from the posterior predictive distribution, and use the
proportion of the ratings that are class 2 as a statistic to summarise each
dataset.

\begin{verbatim}
class2prop <- function() {
  simdata <- posterior_predict(fit_1, anesthesia[, 1:2])
  sum(simdata$rating == 2) / nrow(simdata)
}
ppc_statistics <- replicate(1000, class2prop())
head(ppc_statistics)
\end{verbatim}

\begin{verbatim}
#> [1] 0.5587302 0.4507937 0.4126984 0.4190476 0.4666667 0.2539683
\end{verbatim}

\begin{figure}

{\centering \includegraphics{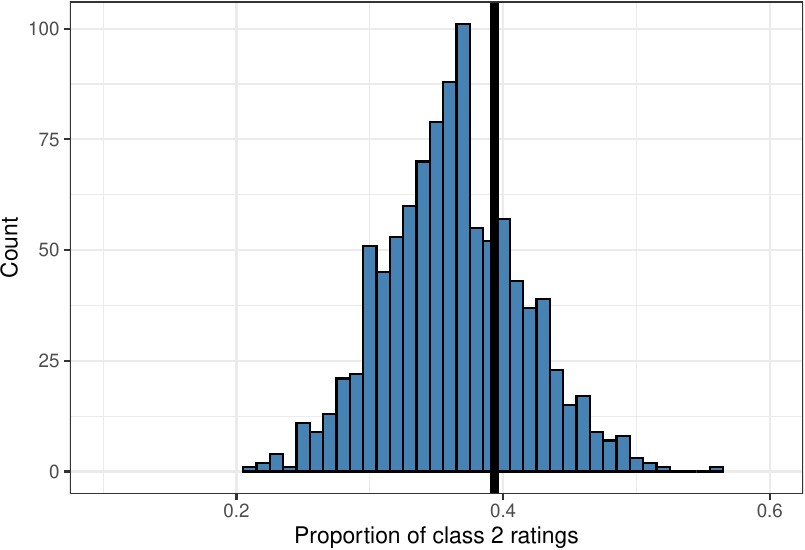} 

}

\caption{An example of a posterior predictive check for the Dawid--Skene model fitted via MCMC to the anaesthesia dataset.  Shown is a histogram of 1,000 simulated datasets from the posterior predictive distribution, each summarised by the proportion of ratings that are class 2.  The vertical black line shows the proportion of ratings that are class 2 in the original dataset, for comparison against the histogram.}\label{fig:plot-ppcs}
\end{figure}

We can then graphically compare the distribution of these statistics to the
value of the same statistic applied to the anaesthesia dataset. The following
commands will create such a plot (shown in \autoref{fig:plot-ppcs}):

\begin{verbatim}
ggplot(data.frame(prop_class_two = ppc_statistics), aes(prop_class_two)) +
  geom_histogram(binwidth = 0.01, fill = "steelblue", colour = "black") +
  geom_vline(xintercept = sum(anesthesia$rating == 2) / nrow(anesthesia),
             colour = "black", linewidth = 2) +
  theme_bw() +
  coord_cartesian(xlim = c(0.1, 0.6)) +
  labs(
    x = "Proportion of class 2 ratings",
    y = "Count"
  )
\end{verbatim}

The statistic calculated on the anaesthesia dataset lies towards the centre of
the distribution of statistics calculated from datasets drawn from the
posterior predictive distribution. This is consistent with the model fitting
the data well. (This is an illustrative example only. A more comprehensive
set of such comparisons should be conducted before concluding that the model
adequately describes the data.)

Second, \CRANpkg{rater} implements the functions \texttt{loo()} and \texttt{waic()}
to provide model comparison metrics using the \CRANpkg{loo} package. The
function \texttt{loo()} calculates an approximation to leave-out-one
cross-validation, a measure of model performance using Pareto Smoothed
Importance Sampling (Vehtari, Gelman, and Gabry 2017). The function \texttt{waic()} calculates
measures related to the Widely Applicable Information Criterion
(Watanabe 2013). Because \CRANpkg{rater} fits models in a Bayesian
framework, information criteria such as the AIC and BIC are not implemented.

In the context of \CRANpkg{rater}, model comparison is only possible between
the different implemented models, not between the same model with different
parameter values or included covariates, as is more commonly the case. For
this reason, considerations of data size and what is known about the
characteristics of the raters should also be taken into account when making
such choices, in addition to the output of \texttt{loo()} and \texttt{waic()}.

To briefly illustrate some of this functionality, we compare the standard
Dawid--Skene model to the class-conditional one.

\begin{verbatim}
loo(fit_1)
\end{verbatim}

\begin{verbatim}
#> 
#> Computed from 4000 by 45 log-likelihood matrix
#> 
#>          Estimate   SE
#> elpd_loo   -233.7 16.9
#> p_loo        19.8  2.6
#> looic       467.5 33.9
#> ------
#> Monte Carlo SE of elpd_loo is 0.1.
#> 
#> Pareto k diagnostic values:
#>                          Count Pct.    Min. n_eff
#> (-Inf, 0.5]   (good)     43    95.6%   734       
#>  (0.5, 0.7]   (ok)        2     4.4%   497       
#>    (0.7, 1]   (bad)       0     0.0%   <NA>      
#>    (1, Inf)   (very bad)  0     0.0%   <NA>      
#> 
#> All Pareto k estimates are ok (k < 0.7).
#> See help('pareto-k-diagnostic') for details.
\end{verbatim}

\begin{verbatim}
loo(diff_model_fit)
\end{verbatim}

\begin{verbatim}
#> 
#> Computed from 4000 by 45 log-likelihood matrix
#> 
#>          Estimate   SE
#> elpd_loo   -245.7 18.1
#> p_loo        10.3  1.2
#> looic       491.3 36.2
#> ------
#> Monte Carlo SE of elpd_loo is 0.1.
#> 
#> All Pareto k estimates are good (k < 0.5).
#> See help('pareto-k-diagnostic') for details.
\end{verbatim}

\begin{verbatim}
loo_compare(loo(fit_1), loo(diff_model_fit))
\end{verbatim}

\begin{verbatim}
#>        elpd_diff se_diff
#> model1   0.0       0.0  
#> model2 -11.9       3.2
\end{verbatim}

From these results, we see that both models fit the data well, with the
class-conditional model being slightly preferred.

\hypertarget{sec:summary}{%
\section{Summary and discussion}\label{sec:summary}}

Rating procedures, in which items are sorted into categories, are subject to both
classification error and uncertainty when the categories themselves are defined
subjectively. Data that results from these types of tasks require a proper
statistical treatment if: (1) the population frequencies of the categories are
to be estimated with low bias, (2) items are to be assigned to these
categories reliably, and (3) the agreement of raters is to be assessed.
To date there have been few options for practitioners seeking software that
implements the required range of statistical models. The R package described in
this paper, \CRANpkg{rater}, provides the ability to fit
Bayesian versions of a variety of statistical models to categorical rating
data, with a range of tools to extract and visualise parameter estimates.

The statistical models that we have presented are based on the Dawid--Skene
model and recent modifications of it including extensions, such as the
hierarchical Dawid--Skene model where the rater-specific parameters are assumed
to be drawn from a population distribution, and simplifications that, for
example, assume exchangeable raters with identical performance characteristics
(homogeneous Dawid--Skene), or homogeneity criteria where classification
probabilities are identical regardless of the true, underlying category of an
item (class-conditional Dawid--Skene).

We provided: (1) an explanation of the type of data formats that categorical
ratings are recorded in, (2) a description of the construction and implementation of the
package, (3) a comparison of our package to other existing packages for fitting
latent class models in R, (4) an introduction to the user interface,
and (5) worked examples of real-world data analysis using \CRANpkg{rater}.

We devoted an entire section to motivating, deriving and explaining the use of
a marginalised version of the joint posterior probability distribution to
remove dependence on the unknown value of the true, underlying rating category
of an item. This is necessary because the No-U-Turn Sampler, the main MCMC algorithm
used in Stan, relies on computing derivatives with respect to all parameters that
must, therefore, be continuously-valued. The technique involves the use of
conditional expectation and is a special case of a more general technique of
conditioning or `Rao--Blackwellization', the process of transforming an
estimator using the Rao--Blackwell theorem to improve its efficiency.

Our package was developed with the classification of medical images in mind,
where there may be a large number of images but
typically only a small number of possible categories and a limited number of
expert raters. The techniques proposed based on extensions of the Dawid--Skene model
readily extend to scenarios where the datasets are much larger, or where the
raters behave erratically (cannot be relied on to rate correctly more
frequently than they rate incorrectly) or are antagonistic (deliberately
attempt to allocate items to a category not favoured by the majority of
raters).

One possible extension of the Dawid--Skene model is to add item- and
rater-specific covariates. These covariates would encode the difficulty of
items and the expertise (or lack thereof) of the raters. This extension is
particularly attractive as it would partially alleviate the strong assumption
of independence conditional on the latent class, replacing it with the weaker
assumption of independence conditional on the latent class and the covariates.
Unfortunately, these types of models would contain many more parameters than
the original Dawid--Skene model making them difficult to fit, especially when
used with relatively small datasets common in the context of rating data in
medical research. Therefore, future methodological research is needed before
these models can be included in the \CRANpkg{rater} package. Latent class
models with covariates can be fitted, although only in a frequentist framework,
using the R package \CRANpkg{poLCA}.

\hypertarget{computational-details}{%
\section{Computational details}\label{computational-details}}

The results in this paper were obtained using
R 4.2.2 (R Core Team 2021)
and the following packages:
\CRANpkg{coda} 0.19.4 (Plummer et al.\ 2006),
\CRANpkg{ggplot2} 3.4.3 (Wickham 2016),
\CRANpkg{knitr} 1.45 (Xie 2014, 2015, 2023),
\CRANpkg{loo} 2.6.0 (Vehtari et al.\ 2023),
\CRANpkg{rater} 1.3.1 (Pullin and Vukcevic 2023),
\CRANpkg{rjtools} 1.0.12 (O'Hara-Wild et al.\ 2023),
\CRANpkg{rmarkdown} 1.0.12 (Xie, Allaire, and Grolemund 2018; Xie, Dervieux, and Riederer 2020; Allaire et al.\ 2023),
\CRANpkg{rstan} 2.26.23 (Stan Development Team 2023).
R itself and all packages used are available from the \href{https://CRAN.R-project.org/}{Comprehensive R Archive
Network (CRAN)}.

\hypertarget{acknowledgments}{%
\section{Acknowledgments}\label{acknowledgments}}

We would like to thank Bob Carpenter for many helpful suggestions, and David
Whitelaw for his flexibility in allowing the first author to work on this paper
while employed at the Australian Institute of Heath and Welfare. Thanks also
to Lars Mølgaard Saxhaug for his code contributions to \CRANpkg{rater}.

\hypertarget{references}{%
\section*{References}\label{references}}
\addcontentsline{toc}{section}{References}

\hypertarget{refs}{}
\begin{CSLReferences}{1}{0}
\leavevmode\vadjust pre{\hypertarget{ref-rmarkdown2023}{}}%
Allaire, JJ, Yihui Xie, Christophe Dervieux, Jonathan McPherson, Javier Luraschi, Kevin Ushey, Aron Atkins, et al.\ 2023. \emph{{rmarkdown}: Dynamic Documents for r}. \url{https://github.com/rstudio/rmarkdown}.

\leavevmode\vadjust pre{\hypertarget{ref-asselineau2018}{}}%
Asselineau, J., A. Paye, E. Bessède, P. Perez, and C. Proust-Lima. 2018. {``Different Latent Class Models Were Used and Evaluated for Assessing the Accuracy of Campylobacter Diagnostic Tests: Overcoming Imperfect Reference Standards?''} \emph{Epidemiology and Infection} 146 (12): 1556--64. \url{https://doi.org/10.1017/S0950268818001723}.

\leavevmode\vadjust pre{\hypertarget{ref-beath2017}{}}%
Beath, Ken J. 2017. {``{randomLCA}: {An} {R} {Package} for {Latent} {Class} with {Random} {Effects} {Analysis}.''} \emph{Journal of Statistical Software} 81 (13): 1--25. \url{https://doi.org/10.18637/jss.v081.i13}.

\leavevmode\vadjust pre{\hypertarget{ref-pyanno}{}}%
Berkes, Pietro, Bob Carpenter, Andrey Rzhetsky, and James Evans. 2011. \url{http://docs.enthought.com/uchicago-pyanno/}.

\leavevmode\vadjust pre{\hypertarget{ref-blackwell1947}{}}%
Blackwell, David. 1947. {``Conditional Expectation and Unbiased Sequential Estimation.''} \emph{Ann. Math. Statist.} 18 (1): 105--10. \url{https://doi.org/10.1214/aoms/1177730497}.

\leavevmode\vadjust pre{\hypertarget{ref-carpenter2017}{}}%
Carpenter, Bob, Andrew Gelman, Matthew D. Hoffman, Daniel Lee, Ben Goodrich, Michael Betancourt, Marcus Brubaker, Jiqiang Guo, Peter Li, and Allen Riddell. 2017. {``Stan: {A} {Probabilistic} {Programming} {Language}.''} \emph{Journal of Statistical Software} 76 (1). \url{https://doi.org/10.18637/jss.v076.i01}.

\leavevmode\vadjust pre{\hypertarget{ref-dawid1979}{}}%
Dawid, A. P., and A. M. Skene. 1979. {``Maximum {Likelihood} {Estimation} of {Observer} {Error}-{Rates} {Using} the {EM} {Algorithm}.''} \emph{Applied Statistics} 28 (1): 20. \url{https://doi.org/10.2307/2346806}.

\leavevmode\vadjust pre{\hypertarget{ref-espeland1989}{}}%
Espeland, Mark A., and Stanley L. Handelman. 1989. {``Using Latent Class Models to Characterize and Assess Relative Error in Discrete Measurements.''} \emph{Biometrics} 45 (2): 587--99. \url{https://doi.org/10.2307/2531499}.

\leavevmode\vadjust pre{\hypertarget{ref-geyer1995}{}}%
Geyer, Charles J. 1995. {``Conditioning in {M}arkov Chain {M}onte {C}arlo.''} \emph{Journal of Computational and Graphical Statistics} 4 (2): 148--54. \url{https://doi.org/10.1080/10618600.1995.10474672}.

\leavevmode\vadjust pre{\hypertarget{ref-goodman1974}{}}%
Goodman, Leo A. 1974. {``Exploratory Latent Structure Analysis Using Both Identifiable and Unidentifiable Models.''} \emph{Biometrika} 61 (2): 215--31. \url{https://doi.org/10.1093/biomet/61.2.215}.

\leavevmode\vadjust pre{\hypertarget{ref-handelman1986}{}}%
Handelman, Stanley L., D. H. Leverett, Mark A. Espeland, and Jennifer A. Curzon. 1986. {``Clinical Radiographic Evaluation of Sealed Carious and Sound Tooth Surfaces.''} \emph{The Journal of the American Dental Association} 113 (5): 751--54. \url{https://doi.org/10.14219/jada.archive.1986.0269}.

\leavevmode\vadjust pre{\hypertarget{ref-JMLR:v15:hoffman14a}{}}%
Hoffman, Matthew D., and Andrew Gelman. 2014. {``The {No}-{U}-{Turn} {Sampler}: Adaptively Setting Path Lengths in {Hamiltonian} {Monte} {Carlo}.''} \emph{Journal of Machine Learning Research} 15: 1593--623. \url{http://jmlr.org/papers/v15/hoffman14a.html}.

\leavevmode\vadjust pre{\hypertarget{ref-hui1980}{}}%
Hui, S. L., and S. D. Walter. 1980. {``Estimating the {Error} {Rates} of {Diagnostic} {Tests}.''} \emph{Biometrics} 36 (1): 167--71. \url{https://doi.org/10.2307/2530508}.

\leavevmode\vadjust pre{\hypertarget{ref-ipeirotis2010}{}}%
Ipeirotis, Panagiotis G, Foster Provost, and Jing Wang. 2010. {``Quality Management on {A}mazon {M}echanical {T}urk.''} In \emph{Proceedings of the ACM SIGKDD Workshop on Human Computation}, 64--67.

\leavevmode\vadjust pre{\hypertarget{ref-jakobsdottir2007}{}}%
Jakobsdottir, Johanna, and Daniel E Weeks. 2007. {``Estimating Prevalence, False-Positive Rate, and False-Negative Rate with Use of Repeated Testing When True Responses Are Unknown.''} \emph{The American Journal of Human Genetics} 81 (5): 1111--13. \url{https://doi.org/10.1086/521582}.

\leavevmode\vadjust pre{\hypertarget{ref-li2014}{}}%
Li, Hongwei, and Bin Yu. 2014. {``Error {Rate} {Bounds} and {Iterative} {Weighted} {Majority} {Voting} for {Crowdsourcing}.''} \emph{arXiv Preprint arXiv:1411.4086}. \url{https://arxiv.org/abs/1411.4086}.

\leavevmode\vadjust pre{\hypertarget{ref-linzer2011}{}}%
Linzer, Drew A., and Jeffrey B. Lewis. 2011. {``\pkg{poLCA}: {An} {R} {Package} for {Polytomous} {Variable} {Latent} {Class} {Analysis}.''} \emph{Journal of Statistical Software} 42 (1): 1--29. \url{https://doi.org/10.18637/jss.v042.i10}.

\leavevmode\vadjust pre{\hypertarget{ref-rjtools}{}}%
O'Hara-Wild, Mitchell, Stephanie Kobakian, H. Sherry Zhang, Di Cook, Simon Urbanek, and Christophe Dervieux. 2023. \emph{{rjtools}: Preparing, Checking, and Submitting Articles to the {``{R Journal}''}}. \url{https://CRAN.R-project.org/package=rjtools}.

\leavevmode\vadjust pre{\hypertarget{ref-mcbook}{}}%
Owen, Art B. 2013. \emph{Monte {Carlo} Theory, Methods and Examples}. \url{https://statweb.stanford.edu/~owen/mc/}.

\leavevmode\vadjust pre{\hypertarget{ref-passonneau2014}{}}%
Passonneau, Rebecca J., and Bob Carpenter. 2014. {``The {Benefits} of a {Model} of {Annotation}.''} \emph{Transactions of the Association for Computational Linguistics} 2 (December): 311--26. \url{https://doi.org/10.1162/tacl_a_00185}.

\leavevmode\vadjust pre{\hypertarget{ref-paun2018}{}}%
Paun, Silviu, Bob Carpenter, Jon Chamberlain, Dirk Hovy, Udo Kruschwitz, and Massimo Poesio. 2018. {``Comparing {Bayesian} {Models} of {Annotation}.''} \emph{Transactions of the Association for Computational Linguistics} 6 (December): 571--85. \url{https://doi.org/10.1162/tacl_a_00040}.

\leavevmode\vadjust pre{\hypertarget{ref-coda}{}}%
Plummer, Martyn, Nicky Best, Kate Cowles, and Karen Vines. 2006. {``CODA: Convergence Diagnosis and Output Analysis for MCMC.''} \emph{R News} 6 (1): 7--11. \url{https://journal.r-project.org/archive/}.

\leavevmode\vadjust pre{\hypertarget{ref-rater}{}}%
Pullin, Jeffrey, and Damjan Vukcevic. 2023. \emph{{rater}: Statistical Models of Repeated Categorical Rating Data}. \url{https://CRAN.R-project.org/package=rater}.

\leavevmode\vadjust pre{\hypertarget{ref-base}{}}%
R Core Team. 2021. \emph{R: A Language and Environment for Statistical Computing}. Vienna, Austria: R Foundation for Statistical Computing. \url{https://www.R-project.org/}.

\leavevmode\vadjust pre{\hypertarget{ref-robert2021raoblackwellization}{}}%
Robert, Christian P., and Gareth O. Roberts. 2021. {``{R}ao-{B}lackwellization in the {MCMC} Era.''} \emph{arXiv:2101.01011 {[}Stat{]}}. \url{https://arxiv.org/abs/2101.01011}.

\leavevmode\vadjust pre{\hypertarget{ref-rzhetsky2009}{}}%
Rzhetsky, Hagit AND Wilbur, Andrey AND Shatkay. 2009. {``How to Get the Most Out of Your Curation Effort.''} \emph{PLOS Computational Biology} 5 (5): 1--13. \url{https://doi.org/10.1371/journal.pcbi.1000391}.

\leavevmode\vadjust pre{\hypertarget{ref-smyth1994}{}}%
Smyth, Padhraic, Usama Fayyad, Michael Burl, Pietro Perona, and Pierre Baldi. 1994. {``Inferring Ground Truth from Subjective Labelling of {V}enus Images.''} In \emph{Advances in Neural Information Processing Systems}, edited by G. Tesauro, D. Touretzky, and T. Leen. Vol. 7. MIT Press. \url{https://proceedings.neurips.cc/paper_files/paper/1994/file/3cef96dcc9b8035d23f69e30bb19218a-Paper.pdf}.

\leavevmode\vadjust pre{\hypertarget{ref-stanguide}{}}%
Stan Development Team. 2021. \emph{Stan {U}ser's {G}uide}. \url{https://mc-stan.org/docs/2_27/stan-users-guide/}.

\leavevmode\vadjust pre{\hypertarget{ref-rstan}{}}%
---------. 2023. {``\pkg{RStan}: The {R} Interface to {Stan}.''} \url{http://mc-stan.org/}.

\leavevmode\vadjust pre{\hypertarget{ref-loo}{}}%
Vehtari, Aki, Jonah Gabry, Mans Magnusson, Yuling Yao, Paul-Christian Bürkner, Topi Paananen, and Andrew Gelman. 2023. {``Loo: Efficient Leave-One-Out Cross-Validation and WAIC for {B}ayesian Models.''} \url{https://mc-stan.org/loo/}.

\leavevmode\vadjust pre{\hypertarget{ref-vehtari2017}{}}%
Vehtari, Aki, Andrew Gelman, and Jonah Gabry. 2017. {``Practical {Bayesian} Model Evaluation Using Leave-One-Out Cross-Validation and {WAIC}.''} \emph{Statistics and Computing} 27 (5): 1413--32. \url{https://doi.org/10.1007/s11222-016-9696-4}.

\leavevmode\vadjust pre{\hypertarget{ref-vehtari2021}{}}%
Vehtari, Aki, Andrew Gelman, Daniel Simpson, Bob Carpenter, and Paul-Christian Bürkner. 2021. {``Rank-{Normalization}, {Folding}, and {Localization}: {An} {Improved} {Rˆ} for {Assessing} {Convergence} of {MCMC} (with {Discussion}).''} \emph{Bayesian Analysis} 16 (2): 667--718. \url{https://doi.org/10.1214/20-BA1221}.

\leavevmode\vadjust pre{\hypertarget{ref-watanabe2013}{}}%
Watanabe, Sumio. 2013. {``A {Widely Applicable Bayesian Information Criterion}.''} \emph{Journal of Machine Learning Research} 14 (March): 867--97.

\leavevmode\vadjust pre{\hypertarget{ref-white2014}{}}%
White, Arthur, and Thomas Brendan Murphy. 2014. {``\pkg{BayesLCA}: {An} {R} {Package} for {Bayesian} {Latent} {Class} {Analysis}.''} \emph{Journal of Statistical Software} 61 (13). \url{https://doi.org/10.18637/jss.v061.i13}.

\leavevmode\vadjust pre{\hypertarget{ref-ggplot2}{}}%
Wickham, Hadley. 2016. \emph{Ggplot2: Elegant Graphics for Data Analysis}. Springer-Verlag New York. \url{https://ggplot2.tidyverse.org}.

\leavevmode\vadjust pre{\hypertarget{ref-knitr2014}{}}%
Xie, Yihui. 2014. {``{knitr}: A Comprehensive Tool for Reproducible Research in {R}.''} In \emph{Implementing Reproducible Computational Research}, edited by Victoria Stodden, Friedrich Leisch, and Roger D. Peng. Chapman; Hall/CRC.

\leavevmode\vadjust pre{\hypertarget{ref-knitr2015}{}}%
---------. 2015. \emph{Dynamic Documents with {R} and Knitr}. 2nd ed. Boca Raton, Florida: Chapman; Hall/CRC. \url{https://yihui.org/knitr/}.

\leavevmode\vadjust pre{\hypertarget{ref-knitr2023}{}}%
---------. 2023. \emph{{knitr}: A General-Purpose Package for Dynamic Report Generation in r}. \url{https://yihui.org/knitr/}.

\leavevmode\vadjust pre{\hypertarget{ref-rmarkdown2018}{}}%
Xie, Yihui, J. J. Allaire, and Garrett Grolemund. 2018. \emph{R Markdown: The Definitive Guide}. Boca Raton, Florida: Chapman; Hall/CRC. \url{https://bookdown.org/yihui/rmarkdown}.

\leavevmode\vadjust pre{\hypertarget{ref-rmarkdown2020}{}}%
Xie, Yihui, Christophe Dervieux, and Emily Riederer. 2020. \emph{R Markdown Cookbook}. Boca Raton, Florida: Chapman; Hall/CRC. \url{https://bookdown.org/yihui/rmarkdown-cookbook}.

\leavevmode\vadjust pre{\hypertarget{ref-yackulic2020}{}}%
Yackulic, Charles B., Michael Dodrill, Maria Dzul, Jamie S. Sanderlin, and Janice A. Reid. 2020. {``A Need for Speed in Bayesian Population Models: A Practical Guide to Marginalizing and Recovering Discrete Latent States.''} \emph{Ecological Applications} 30 (5): e02112. \url{https://doi.org/10.1002/eap.2112}.

\leavevmode\vadjust pre{\hypertarget{ref-zhang2022}{}}%
Zhang, Wen, Jeffrey Pullin, Lyle Gurrin, and Damjan Vukcevic. 2022. {``Investigating the Efficiency of Marginalising over Discrete Parameters in {B}ayesian Computations.''} \emph{arXiv:2204.06313 {[}Stat{]}}. \url{https://doi.org/10.48550/ARXIV.2204.06313}.

\end{CSLReferences}

\address{%
Jeffrey M. Pullin\\
University of Melbourne\\%
School of Mathematics and Statistics\\ Melbourne, Australia\\
\textit{ORCiD: \href{https://orcid.org/0000-0003-3651-5471}{0000-0003-3651-5471}}\\%
}

\address{%
Lyle C. Gurrin\\
University of Melbourne\\%
School of Population and Global Health\\ Melbourne, Australia\\
\textit{ORCiD: \href{https://orcid.org/0000-0001-7052-1969}{0000-0001-7052-1969}}\\%
}

\address{%
Damjan Vukcevic\\
Monash University\\%
Department of Econometrics and Business Statistics\\ Melbourne, Australia\\
\textit{ORCiD: \href{https://orcid.org/0000-0001-7780-9586}{0000-0001-7780-9586}}\\%
\href{mailto:damjan.vukcevic@monash.edu}{\nolinkurl{damjan.vukcevic@monash.edu}}%
}

\end{article}

\end{document}